%% file: main.tex
\pdfoutput=1
\documentclass[
reprint,superscriptaddress,amsmath,amssymb,aps,prd,floatfix]{revtex4-2}
\usepackage[a4paper,left=1.5cm,right=1.5cm,top=3cm,bottom=3cm]{geometry}
\usepackage[toc]{appendix} %add option page to add a cover to the appendices

\usepackage{graphicx}% Include figure files
\usepackage{dcolumn}% Align table columns on decimal point
\usepackage{bm}
\usepackage{amssymb,amsmath,amsfonts,physics}
\usepackage[dvipsnames]{xcolor}
\usepackage{graphicx}
\usepackage{longtable}
\usepackage{verbatim}
\usepackage{color}
\usepackage{mdframed}
\usepackage{soul}
\usepackage{bm}
\usepackage{amsfonts,amssymb,mathrsfs,amsmath,esint}
\usepackage{slashed, cancel}
\usepackage{framed}
\usepackage{mdframed}
\usepackage{simplewick} % For Wick contractions
\allowdisplaybreaks
\usepackage{latexsym}
\usepackage{graphicx}
\graphicspath{{./Figures/}}
\usepackage[dvipsnames]{xcolor}
\usepackage{booktabs}
\usepackage{datetime}
\newdateformat{mydate}{\THEDAY{ }\monthname[\THEMONTH]{ }\THEYEAR}

\usepackage{tikz}
\usepackage{color}
\usepackage{framed}
\usepackage{hyperref}
\usepackage[capitalise]{cleveref}
\hypersetup{colorlinks, citecolor=bluscuro, linkcolor=black, urlcolor=bluscuro}
\definecolor{rossos}{cmyk}{0,1,1,0.55}
\definecolor{bluscuro}{rgb}{0.15, 0.2, .85}
\definecolor{bluchiaro}{cmyk}{1,.3,0.,0.1}

\graphicspath{{./images/}}

\newcommand{\be}{\begin{equation}}
	\newcommand{\ee}{\end{equation}}
\newcommand{\bea}{\begin{eqnarray}}
	\newcommand{\eea}{\end{eqnarray}}
\newcommand{\beq}{\begin{equation}}
	\newcommand{\eeq}{\end{equation}}

\def\beqa{\begin{eqnarray}}

	\def\eeqa{\end{eqnarray}}

\def\lsim{\mathrel{\rlap{\lower4pt\hbox{\hskip0.5pt$\sim$}}
		\raise1pt\hbox{$<$}}}         %less than or approx. symbol
\def\gsim{\mathrel{\rlap{\lower4pt\hbox{\hskip0.5pt$\sim$}}
		\raise1pt\hbox{$>$}}}         %greater than or approx. symbol

\usepackage[normalem]{ulem}
\usepackage{soul}

\begin{document}
	
	\title{Probe of spatial geometry from scalar induced gravitational waves}
	
	\author{Utkarsh Kumar}
	\email{kumarutkarsh641@gmail.com}
	\affiliation{Astrophysics Research Center of the Open University, The Open University of Israel, Ra’anana, Israel}
 \affiliation{Physics Department, Ariel University, Ariel 40700, Israel}
	\author{Udaykrishna Thattarampilly}
	\email{uday7adat@gmail.com}
	\affiliation{Astrophysics Research Center of the Open University, The Open University of Israel, Ra’anana, Israel}
 \affiliation{Physics Department, Ariel University, Ariel 40700, Israel}
	\author{Pankaj Chaturvedi}
	\email{pankaj@phy.nits.ac.in}
	\affiliation{Department of Physics, National Institute of Technology Silchar, Assam, India}
	\date{\today}

	\begin{abstract}
		We investigate a novel probe of spatial geometry of the Universe through the observation of gravitational waves (GWs) induced by first order curvature perturbations. The existence of spatial curvature leaves imprints on the gravitational wave spectrum and formation of primordial black holes. Given the peaked scalar spectrum, the induced spectrum deviates from the flat space power spectrum and the deviation is dependent on the spatial curvature $K$ and reheating temperature $T_{\rm rh}$. For prolonged reheating and negative spatial curvature the spectrum is amplified enough and exhibits an additional peak solely due to $K$ indicating a possible detection by future gravitational wave experiments including LISA and DECIGO. We also observe that the presence of negative spatial curvature improves the constraints on PBH formation, increasing the mass of black holes which are viable dark matter candidates.
  \end{abstract}
	
	\maketitle
	
	\textit{\textbf{Introduction}}--  
 The recent detection of gravitational waves and advancements in gravitational wave detection \cite{LIGOScientific:2017vwq,LIGOScientific:2017ycc,LIGOScientific:2017vox,LIGOScientific:2017bnn,LIGOScientific:2016aoc,LISACosmologyWorkingGroup:2022jok} have opened a new possible avenue for testing models of Cosmology. Early Universe cosmology is notoriously over-reliant on the observations of CMB spectrum which is often insufficient to distinguish between different models \cite{Ben-Dayan:2023rlj,Martin:2013tda,Brandenberger:2020tcr,Brandenberger:2020eyf,Koshelev:2016xqb,Calcagni:2016ofu,Biswas:2014kva,Gasperini:1992em}. Observation of primordial gravitational waves could differentiate or at the very least significantly reduce the number of viable early Universe models and may give further insights into the nature of our Universe. The recent discovery of a stochastic gravitational wave background (SGWB) from pulsar timing array (PTA) experiments, including NANOGrav \cite{NANOGrav:2023gor,NANOGRAV:2018hou}, CPTA \cite{Xu:2023wog}, PPTA \cite{Reardon:2023gzh}, and EPTA+InPTA \cite{EPTA:2023fyk,EPTA:2023xxk}, has prompted extensive research into the possible cosmological origins of the gravitational wave signals observed by these experiments \cite{NANOGrav:2023hvm,Ben-Dayan:2023lwd}. It is not clear if the SGWB signal is of primordial origin, however, several possible Cosmological sources for these GW have been explored including phase transitions\cite{Bringmann:2023opz,NANOGrav:2021flc,Gouttenoire:2023bqy,Chen:2023bms} that may cause bubble collisions, the formation of cosmic strings\cite{Ellis:2020ena,Blasi:2020mfx,Ellis:2023oxs,Maji:2023fhv} or domain walls\cite{Gouttenoire:2023ftk,Zhang:2023nrs}, resonances during reheating, quantum (gravity) fluctuations during inflation (primordial GWs)\cite{Ben-Dayan:2023lwd,Vagnozzi:2023lwo,Benetti:2021uea,Vagnozzi:2022qmc,Das:2023nmm,Datta:2023vbs,Vagnozzi:2023lwo,Vagnozzi:2020gtf,Antoniadis:2022pcn,EPTA:2021crs,Goncharov:2021oub,NANOGrav:2020bcs,NANOGrav:2020bcs,Hobbs:2017oam,Renzini:2022alw,Ben-Dayan:2024aec,Ben-Dayan:2019gll,Kuroyanagi:2014nba,BICEP:2021xfz,Giare:2022wxq}, and GWs induced by large primordial fluctuations, which may also collapse to form primordial black holes (PBHs) \cite{Caprini_2018,Choudhury:2023fjs,Choudhury:2024dzw,Choudhury:2024aji,Choudhury:2024kjj,Datta:2024bqp,Barman:2024ufm,Kohri:2024qpd,Domenech:2020ers,Domenech:2020ssp,Domenech:2021wkk,Domenech:2023fuz,Domenech:2024cjn,Domenech:2024kmh,Balaji:2024hpu,Domenech:2024wao,Calza:2024qxn,RoperPol:2022iel,Ashoorioon:2022raz,Franciolini:2022pav,Franciolini:2022pav,Escriva:2022duf,Li:2023qua,Carr:2023tpt,deJong:2023gsx,Bian:2023dnv,Inomata:2023zup,Figueroa:2023zhu,Gangopadhyay:2023qjr,Andres-Carcasona:2024wqk,Romero-Rodriguez:2024ldc,Kohri:2020qqd,Iovino:2024uxp,Vaskonen:2020lbd,Papanikolaou:2024kjb,He:2024luf}. Among these sources, primordial GWs and those induced by large primordial fluctuations are crucial predictions of inflationary models in cosmology.  The current and future observations naturally raise the question of whether spatial curvature significantly impacts the propagation of GWs in the primordial Universe. Exploring this influence is essential for accurately interpreting observational data and developing theoretical models of the early Universe.

The standard cosmological model encompasses a Universe described by the Friedmann Lemaître Robertson Walker (FLRW) metric with a vanishing spatial curvature. The flatness of the Universe can be explained by invoking the theory of cosmic inflation \cite{Guth:1980zm,Sato:1980yn,Linde:1981mu}. Despite the success of the standard model, recent observations indicate the possibility of a closed Universe \cite{Planck:2018vyg,Handley:2019tkm,DiValentino:2020hov,Efstathiou:2020wem,Park:2018tgj,Wei:2018cov,Denissenya:2018zcv,Ooba:2017ukj,Rana:2016gha,Leonard:2016evk,DiDio:2016ykq,Bull:2013fga,Guth:2012ww,Bel:2022iuf,Kleban:2012ph,Vagnozzi:2020rcz,Vagnozzi:2020dfn} at the $2\sigma$ level.
	
Primordial gravitational waves are much weaker than scalar perturbations ($r\leq 0.03$ \cite{Tristram:2021tvh}), but the first-order scalar perturbations can source scalar-induced gravitational waves (SIGWs), which may dominate at small scales if enhanced, despite typically being smaller at CMB scales. Traditionally, the production of SIGWs has been studied within the framework of a spatially flat cosmological background \cite{Baumann:2007zm,Picard:2023sbz,Domenech:2021ztg,Bari:2023rcw,Balaji:2023ehk,Domenech:2023jve,Chen:2024fir,Domenech:2024rks,Domenech:2024drm,Lozanov:2023aez,Lozanov:2023knf,Lozanov:2023rcd,Adshead:2021hnm,Domenech:2020kqm,Inomata:2019zqy,You:2023rmn,HosseiniMansoori:2023mqh,Gorji:2023sil,Liu:2023pau,Frosina:2023nxu,Choudhury:2023hfm,Harigaya:2023pmw}. The generation and propagation of primordial GWs in a cosmological background with non-vanishing spatial curvature has recently been studied in \cite{DAgostino:2023tgm,Califano:2024tns}. In this work we study second-order tensor perturbations induced by curvature perturbations in presence of non-vanishing spatial curvature (KSIGWs). We demonstrate that spatial curvature has a significant influence on the GWs spectrum. Our results show that inclusion of spatial curvature affects the amplitude of GWs peak depending on the details of the reheating phase.
 
\textit{\textbf{Induced gravitational waves--}} We consider the FLRW metric perturbed up to second order with negligible anisotropic stress, $ ds^2 = a^{2}[-\left(1 + 2 \Phi \right) d\tau^2 + (\left( 1 + 2 \Phi\right)\,\gamma_{ij} + 2 h_{ij} ) dx^{i} dx^{j}] $; here $\tau$ is the conformal time, $a$ is scale factor, $\Phi$, and $h_{ij}$ are the curvature and tensor(transverse-traceless) perturbations respectively. We assume that the Universe is dominated by a perfect fluid with a constant equation of state $w$. Inserting perturbed metric into Einstein field equations we obtain evolution equations for the background and perturbations. Background evolution is determined by the scale factor $a$, obtained by solving the Friedmann equations described in \cite{suppA}. The evolution of second-order tensor perturbations is governed by the following equation:
	\begin{eqnarray}
		h''^{(2)}_{ij} + 2\,\mathcal{H}\,h'^{(2)}_{ij} - \left(\mathcal{D}^{2} - 2\,K \right)\, h^{(2)}_{ij} = 4\,S_{ij}, \label{eq:TensorPert}
	\end{eqnarray}
	where $\mathcal{D}^2 = \gamma^{ij}\,\mathcal{D}_{i}\,\mathcal{D}_{j}$ is the Laplacian operator in spatially curved spacetime. $S_{ij}$ is the source term constituting the first-order scalar-scalar, scalar-tensor, and tensor-tensor contributions. In this work, we are only interested in GWs sourced by first order scalar-scalar correlations and ignore all the other contributions.
 The explicit form of the source term $S_{ij}$ will be given later in this letter. In spatially flat geometry, cosmological perturbations are expanded using the eigenfunctions of the flat-space Laplacian operator. Flat space Laplacian admits plane wave solutions; thus eigen space expansion is the Fourier transform. Laplacian operator for $K \neq 0$ does not admit simple plane wave solutions. Eigen functions for these operators are given by scalar and tensor harmonics \cite{Lindblom:2017maa}. For non-flat geometries, it is convenient to expand the tensor perturbation \( h_{ij} \) in terms of tensor spherical harmonics \( \Upsilon_{ij,lm}^{(n,\lambda)}(\mathbf{y}) \), where \( \mathbf{y} \equiv (r, \theta, \phi) \), such that $ 
	\mathcal{D}^2 \Upsilon_{ij,lm}^{(n,\lambda)} = - (n^2 - 3 K) \Upsilon_{ij,lm}^{(n,\lambda)}.$
	Here, \( n \) represents the wavenumber in spatially curved geometry. Similarly, scalar perturbations $\Phi$ can be expanded in terms of scalar spherical harmonics $\Upsilon_{lm}^{n}$ satisfying $\mathcal{D}^2 \Upsilon_{lm}^{n} = - (n^2 -  K) \Upsilon_{lm}^{n}.$ The spectrum of the scalar and tensor spherical harmonics is complete for any $K$ given $n>0$ \cite{Abbott:1986ct,Hu:1997mn}. For a detailed review of tensor, scalar and vector harmonics the reader may refer to  \cite{suppB}. Expansion of tensor ($h_{ij}$) and scalar ($\Phi$)  perturbations are as follows:  
	\begin{equation}
		h_{ij} = \sum_{nlm\lambda} h_{lm}^{(n,\lambda)} \Upsilon_{ij,lm}^{(n,\lambda)}, \quad \Phi = \sum_{nlm} \Phi^{n}_{lm} \Upsilon_{lm}^{n}.  \label{eq:mode_expansions}
	\end{equation} 
     Similar to the plane wave expansion of perturbations in flat space geometry we can follow the evolution of each mode of perturbations by substituting
	 \cref{eq:mode_expansions} in the evolution equation for perturbations. The equation for evolution of tensor modes obtained this way is 
	\begin{equation}
		h''^{(n,\lambda)}_{lm} + 2\,\mathcal{H}\,h'^{(n,\lambda)}_{lm} + \left(n^{2} - K \right)h^{(n,\lambda)}_{lm}  = 4\,S_{lm}^{ss,(n,\lambda)}\,, \label{eq:mode_h2}
	\end{equation}
	where $h''^{(2),(\lambda)}_{nlm} =  h^{(n,\lambda)}_{lm}$ and $\lambda = +, \times$ stands for  polarization of GWs. The solution of \cref{eq:mode_h2} is given by
	\begin{equation}
		\begin{split}
			h^{(n,\lambda)}_{lm}(\tau) &= \frac{4}{a(\tau)} \int^{\tau} d\bar{\tau}\,G^{n}_{lm}(\tau,\bar{\tau}) a(\bar{\tau}) \,S_{lm}^{ss,(n,\lambda)} \,,
			\label{eq:hper_sol}
		\end{split}
	\end{equation}
 where $G^{n}_{lm}(\tau,\bar{\tau})$ is the Green's function. The source term is derived from the first order super horizon ($n \tau < 1$) modes for scalar perturbations. Using $\Phi^{n}_{lm}(\tau)= \left[3(1+w) / (5+3w)\right]\,\zeta_{lm}^{n}\,T_{lm}^{n}(\tau)$ and form of source term $S^{ss,(n,\lambda)}_{lm}$, mentioned in \cite{suppC}, it is straightforward to rewrite the pure scalar contribution of tensor modes as:
 \begin{equation}
     h^{(n,\lambda)}_{lm}(\tau) = 4\, F(w) \sum_{l'm'}\,\int dn'\,q^{\lambda}\; \zeta^{n'}_{l'm'}\,\zeta^{n-n'}_{lm}\, \mathcal{I}^{n',|n-n'|}_{l'm',lm}(\tau)\,, \label{eq:hk2}
 \end{equation}
 with $F(w) = \left[3(1+w) / (5+3w)\right]^{2}$ and $q^{\lambda} = q_{\lambda}^{ij}\, k_{i}k_{j}$ being the two polarisation of GWs. Note that there is an implicit sum over the two polarisation modes and $\zeta$ is the primordial curvature perturbation. The last term in the integrand of \cref{eq:hk2} is a time integral performed over the retarded time $\bar{\eta}$ on the product of green's function and $f_{l'm',lm}^{n',|n-n'|}$. The details regarding momentum space expansion of the source term $S_{lm}^{ss,(n,\lambda)}$ is explained in \cite{suppC}.  From two-point correlators of the second order tensor perturbations, we can extract the tensor power spectrum induced by scalar perturbations while summing over both polarization states of GWs as
	\begin{equation}
		\sum_{\lambda} \langle h^{(n,\lambda)}_{lm}(\tau) h^{(N,\lambda)}_{LM}(\tau)  \rangle = \prod_{i=n,m,l} \delta^{(3)}(i + I_i)\frac{2 \pi^2}{n^3}\mathcal{P}_{\rm ss}. \label{eq:PPS}
	\end{equation} 
 We employ a change of variables namely $
 x = n\tau, v = n' / n, u = \abs{n - n'} / n,\text{and}\,\gamma_{n} = K / n^{2}$
for further calculations. From \cref{eq:hper_sol,eq:PPS} we can extract the power spectrum for induced gravitational waves as \cite{suppD}
\begin{equation}
    \begin{split}
 \mathcal{P}_{\rm ss}&= F^{2}(w) \int_{0}^{\infty} du \,\int_{\abs{1-u}}^{1+u} \, dv  P_{\zeta}(n v)  P_{\zeta}(n u) \left(1 - \frac{\gamma_{n}}{v^{2}}\right)^{3}  \\& \, \left(1 - \frac{\gamma_{n}}{u^{2}}\right)^{-1}  \,\frac{v^{2}}{u^{2}}\,\left(1 - \frac{(1+v^{2}-u^{2})^{2}}{4 v^{2}}\right)^{2}  \,\Big(\mathcal{I}^{u,v}_{\rm ss}(x)\Big)^{2}\,.
    \end{split} \label{eq:PSS_final}
\end{equation}
The power spectrum for flat case is recovered when $\gamma_{n} \rightarrow 0$.
It is clear from \cref{eq:PSS_final} that the presence of non-zero spatial curvature significantly affects the power spectrum of SIGWs. It is clear that the power spectrum is independent of the indices $l$ and $m$. This is to be expected since the Universe is isotropic. Here onwards in the letter, we will suppress indices $l$ and $m$ and write the cosmological expressions in terms of $x$ and $n$ alone. Thus for example  \(f_{l'm',lm}^{n',|n-n'|}\) will be re written as $f^{u,v}$. 

The kernel $\mathcal{I}^{\rm ss}_{u,v}(x)$, is given by the time integral of Green's function and $f^{u,v}$, ie
\begin{equation}
    \mathcal{I}_{\rm ss}^{u,v}(x) =  \int_{0}^{x}\, d \bar{x} \,f^{u,v}_{\rm ss}(\bar{x},\gamma_{n})\, \frac{a(\bar{x})}{a(x)}\,n\, G^{n}(x,\bar{x}). \label{eq:kernel}
\end{equation}
 Information regarding the source term is contained in 
\begin{equation}    
			f^{u,v}_{\rm ss} = \frac{\alpha + 1}{\left(\alpha + 2\right)\left(\mathcal{H}^2 + K \right)}\left[  \tilde{T}^{v}_{\Phi} \tilde{T}^{u}_{\Phi}\right]  + T_{\Phi}^{u} T_{\Phi}^{v}. \label{eq:fssuv}
\end{equation}
where $\tilde{T^{v}_{\Phi}}= \mathcal{H} T_{\Phi}^{v} + T^{'v}_{\Phi} $ and $\alpha = \left(1 - 3w\right)/ \left(1 + 3w\right)$ with $w $ being the equation of state of the reheating era. Given the energy density of GWs is $\rho_{\rm GW}(\tau)$, the corresponding spectral energy density is given by $\Omega_{\rm GW}(n,\tau) = d \rho_{\rm GW}(\tau) / d \ln n$. The spectral energy density of the SIGWs at the time of reheating is accordingly calculated as:
\begin{align}
		\Omega_{\rm GW,c}(n,\tau_{\rm rh}) &= \frac{\rho_{\rm GW}(n,\tau_{\rm rh})}{3 \mathcal{H}^2\, M_{\rm p}^{2}} = \frac{n^2}{12\, \mathcal{H}^2} \, \overline{\mathcal{P}_{\rm ss}(n,\tau_{\rm rh})}.\label{eq:GW_spectrum}
\end{align}
Finally, the GW spectrum today\cite{Saito:2008jc,Saito:2009jt} is then given by
$\Omega_{\rm GW}(n,\tau_{0})\,h^{2} = \Omega_{\rm rad}(\tau_{0})\,h^{2} \, \Omega_{\rm GW,c}(n,\tau_{\rm rh})$
where $\Omega_{\rm rad}(\tau_{0})\,h^{2} \simeq 4.15 \times 10^{-5}$ is the relative energy density of radiation today \cite{Planck:2018vyg}. We have neglected the detailed dependence on the thermal history of the Universe which can be easily incorporated if required \cite{Saikawa:2018rcs,Kuroyanagi:2008ye,Boyle:2005se,Seto:2003kc,Schwarz:1997gv,Watanabe:2006qe,Weinberg:2003ur}. In order to find the spectral energy density and evaluate the effect of curvature on the spectral energy density, we now calculate the time average of the kernel $\mathcal{I}_{\rm ss}^{u,v}$ and specify the $n$ dependence of $\mathcal{P}_{\zeta}(n)$.

\textit{\textbf{Evaluation of Kernel--}}
The kernel  $\mathcal{I}^{u,v}_{\rm ss}(x)$ contains all the information regarding the time evolution of perturbations. Inclusion of spatial curvature modifies the evolution of scalar and tensor modes and thus the kernel  $\mathcal{I}^{u,v}_{\rm ss}(x)$ is different from that of the flat Universe calculated in \cite{ Domenech:2021ztg,Baumann:2007zm,Picard:2023sbz}. In order to evaluate the Kernel we shall first evaluate the Green's function and scalar perturbations by solving the evolution equations. Unfortunately, both equations do not admit exact analytical solutions, so we solve them in the sub-horizon limit ($x\gg 1$) for a general background fluid. We use the Logolinear series expansion \cite{Lasenby:2003ur,Contaldi:2003zv,Thavanesan:2020lov,Shumaylov:2021qje,Letey:2022hdp} to write down the approximate solution as 
\begin{align}
    G^{n}(x,\bar{x}) &= \frac{\sqrt{x \bar{x}}}{2n}\, \Big(J_{\alpha + 1/2}(\beta_{n} \bar{x}) Y_{\alpha+ 1/2}(\beta_{n} x) \nonumber \\
    &\quad - J_{\alpha+ 1/2}(\beta_{n}x)Y_{\alpha+ 1/2}(\beta_{n}\bar{x})\Big)\,, \label{eq:green_sol}\\
    T^{n}_{\Phi}(y_n) &= (y_{n}/2)^{-\left(\alpha+3/2\right)}\,\Gamma(\alpha+5/2)\, J_{\alpha+3/2}\left(y_{n}\right)\,,
    \label{eq:transfer_sol}
\end{align}
where $J_{\alpha}(x)$ and $Y_{\alpha}(x)$ are the Bessel functions of the first and second kind of order $\alpha$, $c_{s}^{2}$ is sound speed of scalar perturbations and $y_{n}$ is shorthand notation for $c_{s}\, \beta_{n}^{\Phi}\,x$. Functions $\beta_{n}$ and $\beta_{n}^{\Phi}$ encode the effect of spatial curvature and are given by
\begin{equation}
		\beta_{n} = 1 - \frac{\alpha\,\gamma_{n}}{6(1+\alpha)},\quad \beta_{n}^{\Phi} = 1 - \gamma_{n}\,\frac{(1- \alpha\,\left(\alpha -1 \right))}{2\,c_s^2\,\left(\alpha + 1\right)^{2}}\,.
\end{equation}
When the spatial curvature $K$ is zero, so is $\gamma_n$, and functions $\beta_n$ and $\beta_n^{\Phi}$ are one, which yields the familiar flat space solutions. We Insert \cref{eq:green_sol,eq:transfer_sol} in \cref{eq:fssuv} and expand up to leading order of $\gamma_{n}$ to obtain an analytical expression for the source term $f_{\rm ss}^{u,v}$ and $\left(a(\bar{x}) / a(x)\right)\,n\, G^{n}(x,\bar{x})$ as described in \cite{suppE}. The final form of \cref{eq:kernel} is expressed in following form: 
\begin{align}
      \mathcal{I}_{\rm ss}^{u,v}(x) &= \frac{\pi\,2^{2 \alpha + 3} c_{s}^{-2 (\alpha +1)}\,x^{-(\alpha+1/2)}}{2\,\left(\alpha + 2\right)(2\alpha + 3) \left(u v \beta_{u}^{\Phi} \beta_{v}^{\Phi}\right)^{\alpha + \frac{1}{2}}} \,\sum_{i=1}^{4} \Big[ Y_{\alpha + 1/2}\, \nonumber \\& (\beta_{n}x) \mathcal{I}_{J}^{x,i}(u,v)  - J_{\alpha + 1/2} (\beta_{n} x) \,\mathcal{I}_{Y}^{x,i}(u,v) \Big].
\end{align}
Here $\mathcal{I}_{J,Y}^{x,i}(u,v)$ are the integrals over the three Bessel functions described in the supplementary material. For scales that leave sub-horizon ($x\gg 1$) before reheating, we can evaluate these integrals analytically by extrapolating the upper limit of integrals to infinity. Using the analytical expression for integrals obtained in this manner, the kernel can be approximated as:
\begin{align}
   \mathcal{I}_{\rm ss}^{u,v}(x) &= \frac{2^{\alpha+3}\,\left(u \,v \,\beta_{u}^{\Phi} \,\beta_{v}^{\Phi}\, \beta_n \,x\right)^{-(\alpha+1)}}{\pi \, \, c_s^{-3}\,\left(\alpha + 2\right)\,\left(2 \alpha + 3\right)} \, \times \Bigg\{ \frac{\pi}{2} \, Y_{\alpha+ \frac{1}{2}} \left(\beta_{n}\, x\right) \nonumber \\ 
    &  \, \mathcal{I}_{J}(u,v,s)  + J_{\alpha+ \frac{1}{2}}\left(\beta_{n}\, x\right) \, \mathcal{I}_{Y}(u,v,s) \Bigg\}, \label{eq:final_kernel}
\end{align}
where $\mathcal{I}_{J/Y}(u,v,s)$ are
\begin{align}
   & \mathcal{I}_{J/Y}(u,v,s) = \left(1 - \frac{\gamma_{n}}{6(\alpha + 1)}\,x^{2}\right) \tilde{\mathcal{I}}^{1}_{J/Y}(u,v,s) \nonumber \\ 
    &  +  \gamma_{n}\, \frac{2 \,(2 \alpha + 3)}{c_s\, n^{2} \, \pi \, (1+\alpha)^{2}\, (5 \alpha + 9 )\, u v \beta_{u}^{\Phi} \beta_{v}^{\Phi}} \, \tilde{\mathcal{I}}^{2}_{J/Y}(u,v,s)  \nonumber \\ 
    &  +  \gamma_{n} \, \frac{4\,\left(2 \alpha -1 \right)\,(3 + 2\alpha)}{n^{2} \,\, \pi \,c_s \,(1 + \alpha)} \Big[ \, \tilde{\mathcal{I}}^{3}_{J/Y}(u,v,s)  + 2\,\frac{\beta_n}{c_s^{2}} \,  \tilde{\mathcal{I}}^{4}_{J/Y}(u,v,s)\Big]\,, \label{eq:kernel_I}
\end{align} 
where $s$=($(u\beta_{u}^{\Phi})^{2}+(v \beta_{v}^{\Phi})^{2}-\beta_{n}^{2} c_s^{-2}) / 2 u v \beta_{u}^{\Phi}\beta_{v}^{\Phi}$. Both $\mathcal{I}_{J/Y}(u,v,s)$ are expressed in terms of associated Legendre Functions of the first and second kind. We remind the reader that the final expression for the kernel is written by expanding the Bessel functions for large arguments.

\textit{\textbf{Evaluation of the spectrum}}--  Having obtained an expression for the kernel, it is now possible to evaluate \cref{eq:PSS_final} to obtain the spectrum of gravitational waves induced by the primordial scalar fluctuations, for any given power spectrum $\mathcal{P}_{\zeta}(n)$. The primordial power spectrum $\mathcal{P}_{\zeta}(n)$, in general, can be different from the nearly scale-invariance of spectrum observed on the CMB scales. The integral in \cref{eq:PSS_final} can not be evaluated analytically in most cases, however, we are only interested in assessing the effect of curvature on the induced gravitational waves and it is enough to evaluate the spectrum for the simple case of a primordial spectrum with a Dirac Delta-peak situated at the scale $n_{p}$. Let the primordial spectrum be given by 
\begin{align}
    \mathcal{P}_{\zeta}(n) = \mathcal{A}_{\zeta}\, \delta \left(\log (n / n_{p})\right) \label{eq:peaked_spectrum}
\end{align}
where $A_{\zeta}$ is the amplitude of the power spectrum. For our choice of the primordial spectrum it is possible to derive the power spectrum of GWs analytically. Typical examples of inflationary models predicting peaked spectrum are given in \cite{Cai:2020ovp,Inomata:2016rbd,Musco:2018rwt}. We perform the integral and obtain an analytical expression for $\Omega_{\rm GW,c}(x) $, which is related to the induced power spectrum via \cref{eq:GW_spectrum}
\begin{equation}
\begin{split}
    \Omega_{\rm GW,c}(x) &= \frac{\mathcal{A}_{\zeta}^{2}\,F(w)^{2}\,x^{2}\,u_{*}^{2}}{12\,\left(1 + \alpha \right)^{2}}\,\left(1 - \frac{1}{4 u_{*}^{2}}\right)^{2}\left(1 - \frac{\gamma_{n}}{u_{*}^{2}}\right)^{2}\\& \left[1 - \frac{2\,x^{2}\,\gamma_{n}}{3\,\left(1 + \alpha\right)^{2}}\right]\,\mathcal{I}^{2 \, u_{*}, u_{*}}_{\rm ss}(x) \, \Theta(2u_{*}-1). \label{eq:PSS_peak}
\end{split}
\end{equation}
In the above analytical form for $\Omega_{\rm GW,c}(x)$, the spectrum is evaluated at $u=v=n_{p}/n\equiv u_{*}$. The step function in \cref{eq:PSS_peak} ensures the momentum conservation with maximum allowed scale being $2\,n_{p}$.

\begin{figure}[!ht]
\includegraphics[scale=0.50]{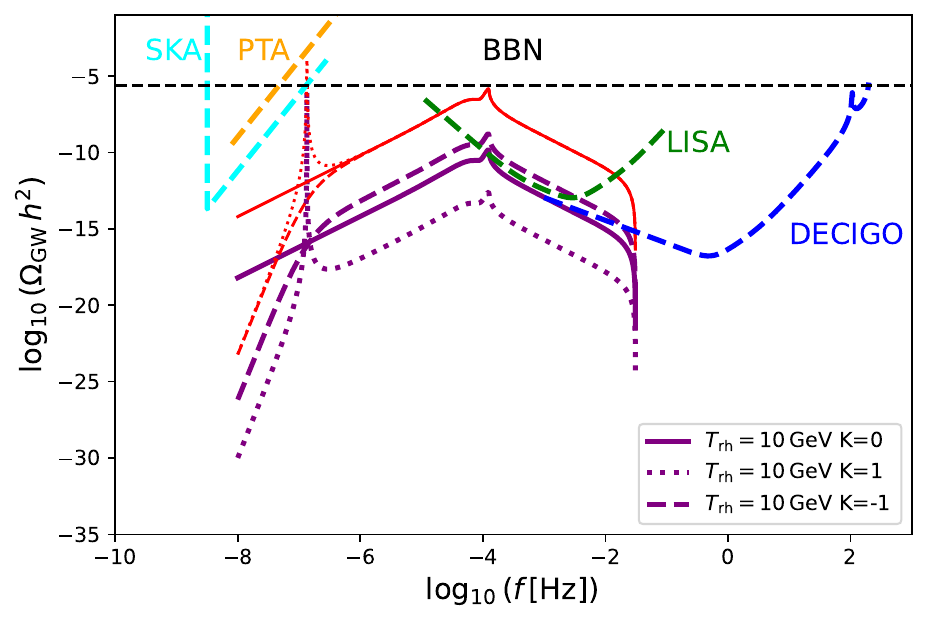}
\includegraphics[scale=0.58]{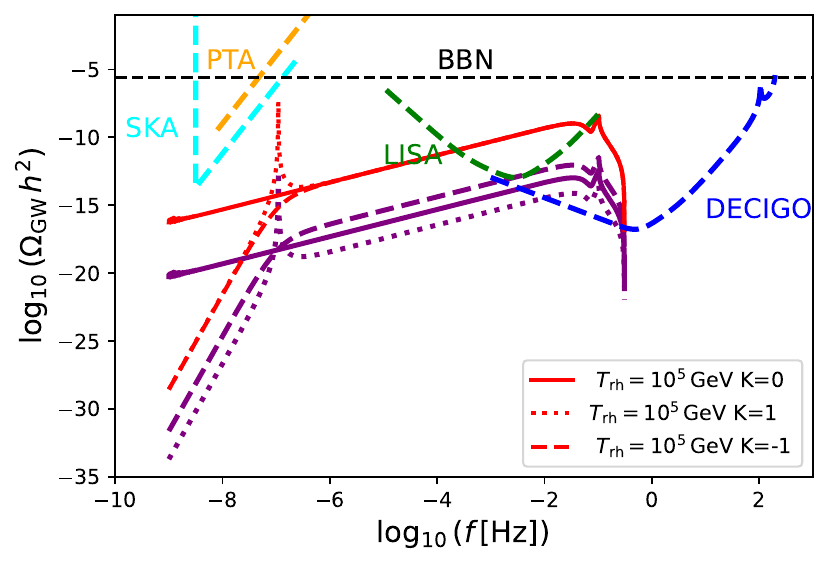}
\caption{GW spectrum induced from the primordial scalar modes in $\rm K = 0$ (solid),$\rm K=+1$ (dotted) and $\rm K=-1$ (dashed) for the Dirac-Delta curvature spectrum. In each panel, we present the spectrum for two values of reheating temperature $\rm T_{\rm rh} = 10 \rm \, GeV$ and $10^{5} \rm \,GeV$ in purple and red respectively. We also include the power-law integrated sensitivity curves \cite{Thrane:2013oya,Schmitz:2020syl} for PTA (\cite{Hazboun:2019vhv}), SKA \cite{Weltman:2018zrl}, LISA (\cite{LISACosmologyWorkingGroup:2022jok}) and DECIGO (\cite{Kawamura:2020pcg}) The horizontal black dashed qualitatively present the current constraint from the BBN \cite{Cooke:2013cba,Aver:2015iza,Cyburt:2004yc,Arbey:2021ysg,Grohs:2023voo}. We choose $n_{p} = 10^{13} \rm Mpc^{-1}$ and $n_{p} = 10^{14} \rm Mpc^{-1}$ with $A_{\zeta} = 2 \times 10^{-2}$ for $\alpha = -0.5$ and $0.5$ in upper and lower panel respectively.}
\label{fig:GWLISA}
\end{figure}
The scalar-induced gravitational wave spectrum $\Omega_{\rm GW}(n,\tau_{0})$ at the present day is shown in \cref{fig:GWLISA}. In both panels of \cref{fig:GWLISA}, we choose the peak of primordial scalar spectrum to be at $n_p = 10^{13} \rm Mpc^{-1}$ and $10^{14} \rm Mpc^{-1}$ such that the peak of the GWs spectrum falls in LISA \cite{LISACosmologyWorkingGroup:2022jok} and DECIGO \cite{Kawamura:2020pcg} sensitivity curves. It is important to note that the GWs spectrum today has several peaks at different frequencies corresponding to $s = \pm 1$. The first peak occurs around $f \sim 10^{-4} \rm Hz$ ($\sim 10^{-1} \rm Hz$) in upper (lower) panel of \cref{fig:GWLISA}. The scale corresponding to the peak can be determined analytically by calculating modes with $s=-1$. The scale corresponding to this value of $s$ is $ n = 2c_s  n_p + \frac{K c^2}{12 c_s n_p}\, \left(18 -12 \alpha ^3-5 \alpha ^2+31 \alpha \right)$ which approximately corresponds to $f \sim 10^{-4} \rm Hz$ and $\sim 10^{-1} \rm Hz$ for $n_p=10^{13}\rm Mpc^{-1}$  and $n_p=10^{14}\rm Mpc^{-1}$ respectively.
The second peak situated at $f \sim 10^{-7} \rm Hz $ corresponds to the other solution, when $s=1$ and the kernel become singular, given by the the frequency mode
   $n = \frac{K\,c^{2}\,\alpha }{12\,c_{s}\,n_{p}\left(1 + \alpha \right)}$ for $K \alpha >0$.
The time evolution of the spectrum across the reheating phase until radiation domination is affected by the spatial curvature as we have observed, and in turn, the amplification due to the spatial curvature depends on the details of the reheating phase. As it is clear from figure \cref{fig:GWLISA} the reheating temperature $T_{\rm rh}$ plays a crucial role in the spectrum of KSIGWs. In \cref{fig:GWLISA}, we show the behavior of $\Omega_{\rm GW}\,h^{2}$ for the two reheating temperature $T_{\rm rh} = 10 \rm GeV$ and $10^{5}\, \rm GeV$ in purple and red respectively for $K =0,+1 \,\text{and}-1$. We note that for the lower values of $T_{\rm rh} \in [10^{-3} \rm GeV, 50 \rm GeV]$ the amplitude of spectrum is enhanced and suppressed for open Universe ($K = -1$) and closed Universe ($K = 1$) respectively compared to a flat Universe. However, for the relatively larger value of reheating temperature, the spectra for open and closed cases coincide with the flat Universe spectrum. A longer reheating phase implies a larger amplification of the spectrum. If we assume that the reheating phase lasts sufficiently long, the induced gravitational waves sourced by the scalar fluctuations are observable by the future GW detectors. These GW observations can, in turn, predict the spatial curvature of the Universe.
Another interesting observation is that the amplitude of the second peak is also affected by $T_{\rm rh}$ and $n_{p}$ and can reach up to SKA and PTA sensitivity curves for $T_{\rm rh} \geq 10^{5} \rm GeV$. 

The behavior of infrared tail of the spectrum for SIGWs becomes important for the non-vanishing $K$. Expanding \cref{eq:PSS_peak} for $(n_{\rm rh} \ll n \ll n_{*})$ we arrive at a general analytical formula for the low-frequency tail of the spectrum. Interestingly enough, there exist several limits in which KSIGW spectrum has distinctive behavior. We illustrate the distinct functional dependence of $\Omega_{\rm GW,c}$ on the scale $n$ for different limits below. 
For $(|K|)^{1/2} \ll n$ 
\begin{align}
\Omega_{\rm GW,c}^{|K|^{1/2} \ll n} \approx \Xi \begin{cases} \left( n / n_{p} \right)^{2(1+\alpha)} & \text{if } \alpha < 0  \\ c_{s}^{4 \alpha}  \left( n / n_{p} \right)^{2(1 - \alpha)} & \text{if } \alpha > 0 \; ,
\end{cases} \label{eq:case1}
\end{align} 
when $(|K|)^{1/2} \sim n$, the spectrum  $\Omega_{\rm GW,c}$ is
\begin{align}
\Xi \begin{cases} c_{s}^{4 \alpha}\,\left(K c^{2} \alpha / n_{p}^{2}\right)^{-2(1+\alpha)}\, \left( n / n_{p} \right)^{2(3+2\alpha)} & \text{if } \alpha < 0  \\ \left(K c^{2} \alpha  / n_{p}^{2} \right)^{-2(1-\alpha)} \left( n / n_{p} \right)^{2(3 - 2\alpha)} & \text{if } \alpha > 0 \;,
\end{cases} \label{eq:case2}
\end{align} 
and finally when $(|K|)^{1/2} \gg n$:
\begin{align}
\Xi \begin{cases} \,c_{s}^{2 \alpha}\,\left(K c^{2} \alpha / n_{p}^{2} \right)^{-2}\, \left( n / n_{p} \right)^{6} & \text{if } \alpha < 0  \\  \,c_{s}^{-2 \alpha}\, \left( K c^{2} \alpha / n_{p}^{2} \right)^{2(2 \alpha -1 )} \left( n / n_{p} \right)^{2(3 - 2\alpha)} & \text{if } \alpha > 0 \; ,
\end{cases} \label{eq:case3}
\end{align} 
where
\begin{equation}
   \Xi = A_{\zeta}^{2}\, F(\alpha)^{2}\,\, c_{s}^{-(4 \alpha + 6)}\left(1 -  \frac{2 \,c^{2} K} {3 \,n_{\rm rh}^{2}} \right) \left[\frac{n_{p}}{n_{\rm rh}}\right]^{2 \alpha}\, .
\end{equation}
It is evident from \cref{eq:case1,eq:case2,eq:case3} that GW spectrum reduces to flat geometry case whenever $n_{r}^{2} \gg |K|$.
\begin{figure}[!ht]
\includegraphics[scale=0.65]{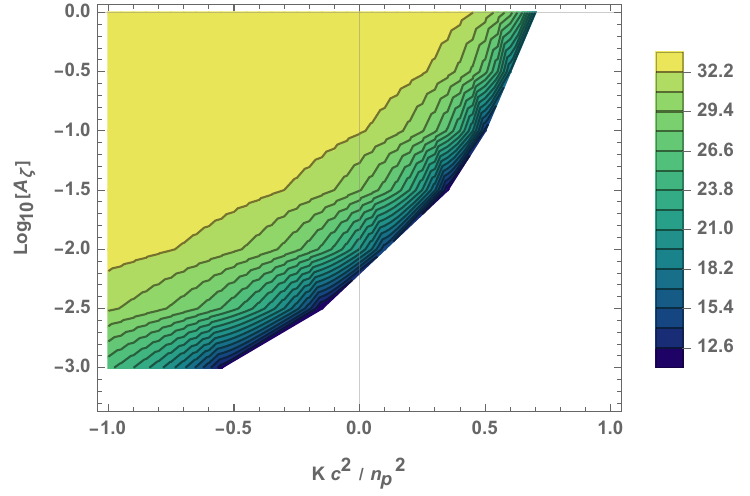}
\includegraphics[scale=0.65]{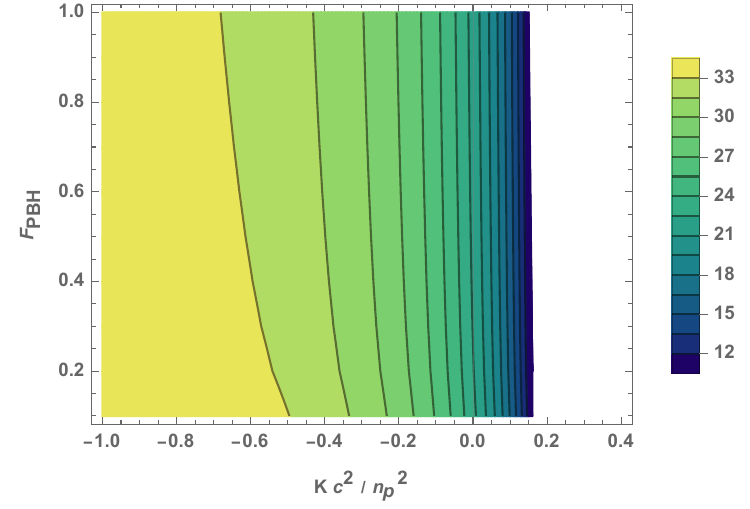}
\caption{\textit{Upper}: The region of parameter space for $A_{\zeta}$ and $ \log_{10}M_{\rm PBH}(\rm g)$ such that PBHs comprise the totality of DM i.e. $F_{\rm PBH} = 1$. \textit{Lower}: Parameter space for $F_{\rm PBH}$ and $\log_{10}M_{\rm PBH}(\rm g)$ for given $A_{\zeta} = 10^{-2}$. We choose $\alpha = - 0.5$ and $T_{\rm rh} = 10^{4} \rm GeV$. for both panels.}
\label{fig:fpbh=1}
\end{figure}

\textit{\textbf{PBH Formation}}--  So far, we have studied the GW signature of a peaked primordial spectrum in a Universe with spatial curvature.  
The amplitude of GW spectral density measured today will be determined by the peak amplitude of the spectrum. If the peak value, $A_{\zeta}$ is large enough, density fluctuations in the early Universe could collapse and generate PBHs. The mass of PBHs $M_{\rm PBH}$ is related to the peak frequency of the gravitational wave spectrum by the relation (\cite{Alabidi:2013lya,Domenech:2019quo}) 
\begin{equation}
    f_{\rm GW} \approx 7.08 \times 10^{-6} \rm Hz \left(\frac{T_{\rm rh}}{10^{2} \rm GeV}\right)^{\frac{\alpha}{\alpha +2}} \left(\frac{1.8 \times 10^{27} \rm g}{M_{\rm PBH}}\right)^{\frac{1}{\alpha + 2}}. \label{eq:GW_freq}
\end{equation}
The candidacy of PBHs as dark matter can be tested by estimating the fraction of PBH to dark matter ($F_{\rm PBH}$) given by \cite{Alabidi:2013lya,Carr:2009jm,Young:2019osy,Sasaki:2018dmp}
\begin{equation}
    F_{\rm PBH} \approx 2 \beta \times 10^{12} \left(\frac{f_{\rm GW}}{2.6 \times 10^{-5} \rm Hz}\right)^{1-\alpha} \left(\frac{T_{\rm rh}}{10^{2} \rm GeV}\right)^{\alpha} \label{eq:F_pbh}.
\end{equation}
%We can constrain the spatial curvature on small scales by the abundance of PBHs.
The PBH mass fraction at the formation ($\beta$) evaluated using the Press–Schechter formalism \cite{Escriva:2019phb,DeLuca:2020ioi,Sureda:2020vgi} is $\frac{1}{2}\, \rm erfc\left(\delta_{c} / \sqrt{2}\,\sigma_{\delta}\right)$. In the presence of spatial curvature, we evaluate $\sigma_{\delta}$ and $\delta_{c}$ as
\begin{equation}
    \delta_{c} = F(\alpha)^{\frac{1}{2}} \sin^{2}\left[\frac{\pi}{2}\sqrt{\frac{1-\alpha^{2}}{3}}\right],\sigma_{\delta} =(F(\alpha) A_{\zeta})^{\frac{1}{2}} \left[1-\frac{K c^{2}}{n_{p}^{2}}\right]^{2}.
\end{equation}
To constrain the parameters of the GW spectrum, we use the fact that PBH density cannot exceed that of dark matter i.e. $1 \geq F_{\rm PBH}$. When $F_{\rm PBH} \simeq 1$ most of the dark matter in the Universe is accounted for by the existence of PBH. A lower fraction would mean that dark matter comprises several different matter contents that only interact via gravity, a significant fraction of which are PBHs. 
This inequality together with \cref{eq:GW_freq,eq:F_pbh} constrains the parameter space for PBHs, resulting in allowed values of $A_{\zeta}$ and $M_{\rm PBH}$ as depicted in \cref{fig:fpbh=1}. The upper panel in \cref{fig:fpbh=1} shows possible values of $A_{\zeta}$, $\rm M_{\rm PBH}$, and  $K/n_p^2$ for which dark matter is entirely comprised of PBHs. As it is clear from the figure the presence of spatial curvature increases the volume of the allowed parameter space, improving the possibility of this scenario. The lower panel of \cref{fig:fpbh=1} shows possible values of  $\rm M_{\rm PBH}$ and  $K/n_p^2$ when we loosen the constraint to allow for only a fraction of dark matter being comprised of PBHs. We have fixed $A_{\zeta}$ to be $10^{-2}$ in this case. A negative spatial curvature allows for the existence of black holes with larger masses than those anticipated from a flat Universe.

 \textit{\textbf{Conclusions}}-- We studied the
gravitational waves induced by first-order curvature perturbations and PBH formation in a Universe with non-zero spatial curvature. We have derived the power spectrum of second-order tensor perturbations induced by a scalar source term in a Universe with spatial curvature, which is absent in the current literature to the best of our knowledge. For the Dirac-Delta scalar spectrum, our estimations show that the amplitude of GWs depends upon the geometry of the Universe and the details of the reheating phase. Specifically, We observe a significant enhancement in the amplitude of GWs when the spatial curvature assumes its value to be negative for the reheating temperature, $T_{\rm rh} \in [10^{-3} \rm  GeV, 50  \rm GeV]$. Furthermore, due to a non-vanishing spatial curvature, the resulting GWs spectrum has additional peaks which is the smoking gun feature of KSIGWs. The main features of KSIGWs are observable for a wide range of frequencies varying from PTA to LISA scales and for different reheating phases with low and high $T_{\rm rh}$.

In this letter, we have only considered a Dirac Delta spectrum which serves the purpose of showing the main features of KSIGWs. However, scale-invariant and varying peak width spectrum may be considered for generating GW spectrum in curved FLRW background. We stress that the consideration of another spectrum may result in the shape of GW spectrum being different from the flat case. However, the features of the spectrum will remain similar to those discussed here, for the Dirac Delta spectrum. We note that the scale-dependent curvature corrections to the induced tensor spectrum are responsible for the shape of the GW spectrum. We have also derived the constraints on the PBH abundances. The presence of spatial curvature significantly improves the volume of the parameter space for PBHs to be a viable dark matter candidate. A negative spatial curvature will allow for the existence of massive primordial black holes, shown to be unfeasible candidates if the Universe is flat. We leave the analysis of KSIGWs for recent PTAs signals for different choices of scalar spectrum for our future endeavors. 

\textit{\textbf{Acknowledgments}}-- We thank Prof. Ido Ben-Dayan for his valuable comments. We also thank Amresh Verma and Pritha Bari for helpful discussions. U.K. and U.T. are supported in part by the “Program of Support of High Energy Physics” Grant by Israeli Council for Higher Education. We also acknowledge the Ariel HPC Center at Ariel University for providing computing resources that have contributed to the research results reported in this paper.

	\newpage
	
	\bibliography{ref.bib}
	
	\newpage
	\pagebreak
	\appendix
	\onecolumngrid
	\input{supplement.tex}

\end{document}

%% file: supplement.tex
% Begin of SM

%TC:ignore
\widetext
\begin{center}
\textbf{\Large Supplementary Material}
\end{center}
%%%%%%%%%%%%%%%%%%%%%%%%%%%%%%%%%%%%%%%%%%%%%%%%%%%%%%%%%%%%%%%%%%%%%%%%%%%%%
\section{Perturbations and evolution equations in the presence of spatial curvature} \label{sec:appendixA}
We write and expand Einstein equations for a maximally symmetric spacetime assuming non-zero spatial curvature. The first-order and second-order perturbation equations, calculated in the Newtonian gauge, were derived using the \textit{xPand} package in \textit{Mathematica}. An isotropic and homogeneous Universe with a non-zero spatial curvature is described by the Friedmann–Lemaître–Robertson–Walker (FLRW) metric given by:
\begin{eqnarray}
    ds^{2} \equiv g_{\mu \nu} \, dx^{\mu}\,dx^{\nu} = a(\tau)^{2} \, \left[-d \tau^{2} + \gamma_{ij}\,dx^{i}\,dx^{j}\right] ,\label{eq:FRW}
\end{eqnarray}  
where the scale factor $a(\tau)$ is a function of conformal the time and $\gamma_{ij}$ is the maximally symmetric metric of spatial hyper-surfaces given by:
\begin{eqnarray}
    \gamma_{ij}\,dx^{i}\,dx^{j} = \frac{dr^{2}}{1- K\,r^{2}} + r^{2}\,\left( d \theta^{2} + \sin^{2} \theta \, d\phi^{2}\right),\label{eq:SFRW}
\end{eqnarray}
\( K \) is the spatial curvature of the Universe. $K$ can take on different values: \( K = 0 \) corresponds to a flat universe, \( K > 0 \) denotes a closed universe, and \( K < 0 \) indicates an open universe. In the presence of an adiabatic perfect cosmic fluid with the stress-energy tensor given by
\begin{equation}
T_{\mu\nu}= (\rho+p)v_{\mu}v_{\nu}+p g_{\mu\nu},\label{eq:Tmn}
\end{equation}
the Einstein field equations for the spacetime in eq.(\ref{eq:FRW}) can be expressed as 
\begin{equation}
\mathcal{H}^2= \frac{1}{3} \rho  \, a^{2} - K, \qquad \text{and} \qquad \mathcal{H}' + \mathcal{H}^{2} = \frac{1}{6}\, \left(\rho - 3 p\right)\, a^{2}-K,\label{eq:EEA}
\end{equation} 
where $\rho$ and $p$ are, respectively the energy density and the pressure, $v^{\mu}$ is the four velocity, and $\mathcal{H} \equiv a' / a$ is the conformal Hubble parameter. The scale factor can be determined by solving the eqs. (\ref{eq:EEA}), as 
\begin{eqnarray}
    a(\tau) &\propto & 
    \begin{cases}
     \, \sinh{\left(\frac{1 + 3w}{2}\, \sqrt{|K|}\, \tau \right)}^{2/(1+3w)} & \text{if }\, K<0 \\
    \tau^{2/(1+3w)} & \text{if } \, K=0 \\
     \, \sin{\left(\frac{1 + 3w}{2}\, \sqrt{K}\, \tau \right)}^{2/(1+3w)} & \text{if }\, K>0 
    \end{cases}.\label{eq:aform}
\end{eqnarray}

Given the background eq. (\ref{eq:FRW}) in the conformal Newtonian gauge, we aim to derive the evolution equations for cosmological perturbations up to second order. The line element for cosmological perturbations is given by 
\begin{eqnarray}
    ds^2 &=& a^{2}(\tau) \left[-\left(1 + 2 \Phi + \Phi^{\left(2\right)}\right)\,d\tau^2 + \left(\left( 1 + 2 \Psi + \Psi^{\left(2\right)}\right)\,\gamma_{ij}  + 2 h_{ij} + h_{ij}^{\left(2\right)}\right) \, dx^{i} dx^{j}\right],
\end{eqnarray}
where metric perturbations $\Phi^{\left(n\right)}$, $\Psi^{\left(n\right)}$, $h_{ij}^{\left(n\right)}$ are lapse, curvature and tensor perturbations. We do not include vector perturbations at any given order, as they are diluted away during the inflationary phase, even in spatially non-flat cases.  The tensor perturbations $h_{ij}$ are transverse-traceless (TT) satisfying the $\partial^{i}\,h_{ij}^{\left(n\right)} = \delta^{ij}\,h_{ij}^{\left(n\right)} $, where $\left(n\right)$ is the order of perturbations. 

\begin{paragraph}{\textbf{First order scalar and tensor perturbations}:} The evolution equations for first-order cosmological perturbations can be derived by explicitly writing down the perturbed Einstein equations to the first order, for scalar and tensor modes. Equations for the $00$, $0i$, and $ii$ components for the first order scalar perturbation are given by,
\begin{eqnarray}
  & & 3 K\,\Phi - 3\,\mathcal{H}\, \Phi' +  \, \mathcal{D}_{\alpha}\,\mathcal{D}^{\alpha}\, \Phi  = a^{2} \left(\frac{1}{2} \, \delta \rho + \rho \, \Phi  \right) \label{eq:scalar00}\\ 
  & & \mathcal{H}\,\Phi + \Phi' =  \frac{a^{2}\, \rho }{3\left(\mathcal{H}^{2} + K \right)} \, \left(\mathcal{H}\,\Phi + \Phi'\right) \label{eq:scalar0i}\\
 & &  \Phi'' + 3\mathcal{H}\,\Phi' + \left( 2 \, \mathcal{H}' + \mathcal{H}^2 -K \right)\, \Phi = \frac{1}{2} \delta\, . \label{eq:scalarii}
\end{eqnarray}
We can combine the equations (\ref{eq:scalar00}) and (\ref{eq:scalarii}) to write the evolution equation for $\Phi$ as
\begin{eqnarray}
    \Phi'' + 3 \, \mathcal{H}\left(1 + c_{s}^{2} \right)\,\Phi' + \left(2\,\mathcal{H}' + \left(1 + 3\, c_{s}^{2}\right)\,\left(\mathcal{H}^{2} -K\right)\right)\,\Phi  - c_{s}^{2}\, \mathcal{D}^{2}\, \Phi= 0,\label{eq:first_scalar_1}
\end{eqnarray}
where $c_{s}^2 = \delta\, p / \delta\, \rho$ is the sound speed of scalar perturbations. For the first-order tensor evolution equation, only transverse-traceless (TT) quantities contribute to the perturbed Einstein equations. The TT terms can be extracted by applying the following TT projection operator $\Lambda^{ij}_{ab}$ to first-order Einstein Equations
\begin{eqnarray}
    \Lambda^{ij}_{ab} &=& \left(\gamma^{i}_{a} - \frac{\mathcal{D}^{i}\mathcal{D}_{a}}{\mathcal{D}^{2}}\right)\,\left(\gamma^{j}_{b} - \frac{\mathcal{D}^{j}\mathcal{D}_{b}} {\mathcal{D}^{2}}\right) - \frac{1}{2}\, \left(\gamma_{ab} - \frac{\mathcal{D}_{a}\mathcal{D}_{b}}{\mathcal{D}^{2}}\right)\,\left(\gamma^{ij} - \frac{\mathcal{D}^{i}\mathcal{D}^{j}}{\mathcal{D}^{2}}\right),\label{eq:TTpro}.
\end{eqnarray}
Applying the TT gauge we can derive the evolution equation for the first-order tensor perturbations to be:
\begin{eqnarray}
    h_{\mu\nu}'' + 2\,\mathcal{H}\,h_{\mu\nu}' - \left(\mathcal{D}^{2} - 2\,K \right)\, h_{\mu\nu} = 0.
\end{eqnarray}
    
\end{paragraph} 

\begin{paragraph}{\textbf{Second order scalar and tensor perturbations}:}
For induced gravitational waves, the relevant evolution equations correspond to the second-order tensor perturbations, derived from the spatial components of the second-order Einstein equations. %To determine the evolution equations we begin by considering the
Expressions for the different components of the second-order Einstein tensor $G^{(2)}$ are given by:
\begin{equation}
    \begin{split}
        G_{00}^{(2)} &= - \, h'_{\alpha \beta} h'^{\alpha \beta}- 8 \,h'_{\alpha \beta} h'^{\alpha \beta}\, \mathcal{H} - 4 \,h'_{\alpha \beta} h'^{\alpha \beta} \, K + 48\, K \, \Phi^2 - 24 \, \mathcal{H}\, \Phi\,\Phi' +6\, \Phi'^{2}+ 6\, K\, \Phi^{(2)} \\& + 6\, K\, \Psi^{(2)} 
        -6\, \mathcal{H}\, \Psi'^{(2)} + 24\, \Phi\, \mathcal{D}^{2} \Phi + 2\,\mathcal{D}^2\, \Psi^{(2)} + 6\, \mathcal{D}_{\alpha} \Phi\, \mathcal{D}^{\alpha} \Phi - 4\, h^{\alpha \beta}\, \mathcal{D}_{\beta} \mathcal{D}_{\alpha} \, \Phi + 4\, h^{\alpha \beta}\, \mathcal{D}^{2} \, h_{\alpha \beta} \\& - 2\, \left(\mathcal{D}_{\beta} h_{\alpha \gamma} \right)\,\left(\mathcal{D}^{\gamma} h^{\alpha \beta} \right) + 3\,\left(\mathcal{D}^{\gamma} h^{\alpha \beta} \right)\,\left(\mathcal{D}_{\gamma} h_{\alpha \beta} \right)
    \end{split}
\end{equation}

\begin{equation}
    \begin{split}
        G_{i0}^{(2)} &= -4\, h'_{i \alpha} \, \mathcal{D}^{\alpha} \Phi + 4\,h_{i \alpha}\, \mathcal{D}^{\alpha} \Phi' - 4 h^{\alpha \beta} \, \mathcal{D}_{\beta} h'_{i \alpha} + 2\, h^{\alpha \beta}\, \mathcal{D}_{i} h_{\alpha \beta}  + 4 \, h^{\alpha \beta} \, \mathcal{D}_{i} h'^{\alpha \beta}   - 8\,\mathcal{H}\, \Phi \, \mathcal{D}_{i} \Phi  + 4\, \Phi' \, \mathcal{D}_{i}\, \Phi \\& + 8 \, \Phi\, \mathcal{D}_{i} \Phi' + 2 \mathcal{H}\, \mathcal{D}_{i} \Phi^{(2)} + 2 \mathcal{H}\, \mathcal{D}_{i} \Psi^{(2)} 
    \end{split}
\end{equation}

\begin{equation}
	\begin{split}
		G_{ij} &= -4\, h'^{\alpha}_{i} \, h'_{j \alpha} + h''^{(2)}_{i j} + 3\, \gamma_{i j}\, h'_{\alpha \beta} \,h'^{\alpha \beta} + 4\, \gamma_{i j} \, h''_{\alpha \beta} \,h^{\alpha \beta} + 2\, \mathcal{H}\, h'_{i j} + 8 \,\gamma_{ij}\, h^{\alpha \beta} \,h'_{\alpha \beta} - 2 \, \mathcal{H}^{2}\, h_{i j}^{(2)} - 4\, h_{i j}^{(2)}\, \mathcal{H}'  \\& + 4 \,K\, \gamma_{ij} \, h_{\alpha \beta} \, h^{\alpha \beta} - 4\, h''_{i j} \, \Phi - 8 \, \mathcal{H}\, h_{ij}' \, \Phi + 8\,\mathcal{H}^2 \, h_{ij}\, \Phi + 16\, \mathcal{H}'\,\Phi -16 \, \gamma_{i j}\, \mathcal{H}^2\,\Phi^{2}  -32 \, \gamma_{i j} \, \mathcal{H}'\, \Phi^{2}  + 32\, \mathcal{H}\,h_{i j} \, \Phi'  \\& - 40 \, \gamma_{i j} \, \mathcal{H}\, \Phi \, \Phi' - 2 \gamma_{i j} \Phi'^{2} + 12 \, h_{i j}\,  \Phi'' - 8 \, \gamma_{i j}\, \Phi\,\Phi'' + 2\gamma_{i j}\, \mathcal{H}^2\,\Phi^{(2)}  + 4\, \gamma_{i j}\,\mathcal{H}'\, \Phi^{(2)} + 2\, \gamma_{i j}\, \mathcal{H}\, \Phi'^{(2)}+ 2\, \gamma_{i j}\, \mathcal{H}^{2}\, \Psi^{(2)} \\& + 4 \gamma_{ij} \, \mathcal{H}'\, \Psi^{(2)} + 4\, \gamma_{i j}\, \mathcal{H}\, \Psi'^{(2)} + 2\, \gamma_{ij}\, \Psi''^{(2)} - 4\, \Phi\, \mathcal{D}^{2}\, h_{ij} - \mathcal{D}^{2}\, h^{(2)}_{ij}   -4 h_{ij} \, \mathcal{D}^{2}\, \Phi -8\,\gamma_{ij}\,\Phi\,\mathcal{D}^{2}\Phi + \, \gamma_{i j} \, \mathcal{D}^{2}\,\Phi^{(2)} \\& -\, \gamma_{i j} \, \mathcal{D}^{2}\,\Psi^{(2)} + 4 \,h_{j}^{\alpha} \, \mathcal{D}_{\alpha}\mathcal{D}_{i} \, \Phi + 4 \,h_{i}^{\alpha} \, \mathcal{D}_{\alpha}\mathcal{D}_{j} \, \Phi - 8\,\mathcal{D}_{\alpha} h_{ij}\,\mathcal{D}^{\alpha} \Phi - 6\gamma_{ij}\, \mathcal{D}_{\alpha} \Phi \, \mathcal{D}^{\alpha} \Phi + 4\, h^{\alpha \beta} \, \mathcal{D}_{\beta}\mathcal{D}_{\alpha}\, h_{ij} \\& - 4 \gamma_{i j}\,
		h^{\alpha \beta} \, \mathcal{D}_{\beta}\mathcal{D}_{\alpha}\, \Phi - 4\,h^{\alpha\beta} \, \mathcal{D}_{\beta} \mathcal{D}_{i} \, h_{j \alpha} - 4\,h^{\alpha\beta} \, \mathcal{D}_{\beta} \mathcal{D}_{j} \, h_{i \alpha} - 4 \mathcal{D}_{\alpha} \, h_{j \beta} \mathcal{D}^{\beta} \, h_{i}^{\alpha} + 4 \mathcal{D}_{\beta} \, h_{j \alpha} \mathcal{D}^{\beta} \, h_{i}^{\alpha} + 4\, \gamma_{i j} \mathcal{D}_{\gamma} h_{\alpha \beta}\,\mathcal{D}^{\gamma} h^{\alpha \beta} \\& +4 \, \mathcal{D}^{\alpha} \Phi \, \mathcal{D}_{i} h_{j \alpha} + 2 \mathcal{D}_{i}\, h^{\alpha \beta} \, \mathcal{D}_{j}\, h_{\alpha \beta} + 4 \mathcal{D}^{\alpha} \Phi \, \mathcal{D}_{j} h_{i \alpha} +  4 \mathcal{D}_{i} \Phi \, \mathcal{D}_{j} \Phi + 4 h^{\alpha \beta}\, \mathcal{D}_{j}\mathcal{D}_{i}\, h_{\alpha \beta} + 8 \Phi\, \mathcal{D}_{j}\,\mathcal{D}_{i} \Phi - \mathcal{D}_{j}\,\mathcal{D}_{i}\, \Psi^{(2)}.
  \end{split}
  \end{equation}
Similarly, the second-order perturbations of the energy-momentum tensor are:
\begin{eqnarray}
        T_{00}^{(2)} &=& a^4\left(\, \rho^{(2)} + 4\, \rho^{(1)} \Phi + 2\,\, \rho \, \Phi^{(2)} + 2 \, \left(\rho + P\right)\, \mathcal{D}_{\alpha} v \mathcal{D}^{\alpha} v\right)   \\
       T_{i0}^{(2)}  &=& -a^4\,v^{(2)}_{i}\,\left(P + \rho\right) - 4 a^{4}\, h_{0\alpha}\, \left(P + \rho \right)\, \mathcal{D}^{\alpha} v \\
       T_{ii}^{(2)} &=& a^2\left(p^{(2)}-4p^{(1)}\Phi^{(1)}-2p\,\Phi^{(2)}+\frac{2}{3} \, \left(\rho + P\right)\, \mathcal{D}_{\alpha} v \mathcal{D}^{\alpha} v\right).
\end{eqnarray}
Applying the TT projection operator defined in eq.(\ref{eq:TTpro}) on the second-order Einstein equations written  using he second-order expansions of Stress-Energy and Einstein tensors, one can obtain the following evolution equations for the second-order tensor perturbations, 
\begin{eqnarray}
    h_{ij}^{(2)''} + 2\,\mathcal{H}\,h_{ij}^{(2)'} - \left(\mathcal{D}^{2} - 2\,K \right)\, h_{ij}^{(2)} = S_{ij},\label{eq:hij2}
\end{eqnarray}
where, the tensor $S_{ij}$ acts as a source term for induced waves. The source term can be decomposed into three components: quadratic contributions from first-order scalar perturbations (ss), quadratic contributions from first-order tensor perturbations (tt), and a scalar-tensor cross term (st) :
\begin{eqnarray}
   S_{ij} = S^{ss}_{ij}+S^{st}_{ij}+S^{tt}_{ij}.
\end{eqnarray}
 The scalar-scalar source term $S^{ss}_{ij}$ is given by
\begin{eqnarray}
  S^{ss}_{ij} &=&\frac{8}{3(1+w)(\mathcal{H}^2+K)} \left(\partial_{i} (\mathcal{H}\Phi+\Phi')\partial_{j} (\mathcal{H}\Phi+\Phi') \right)+4\partial_{i}\Phi\partial_{j}\Phi .\label{eq:sourceSS_1}
\end{eqnarray}
If we neglect the contributions from the $S^{st}_{ij}$ and $S^{tt}_{ij}$ as they are subdominant compared to the scalar contributions, the evolution equations in eq.(\ref{eq:hij2}) can be written as
\begin{eqnarray}
    h_{ij}^{(2)''} + 2\,\mathcal{H}\,h_{ij}^{(2)'} - \left(\mathcal{D}^{2} - 2\,K \right)\, h_{ij}^{(2)} = S^{ss}_{ij}.\label{eq:hij2f}
\end{eqnarray}

\end{paragraph} 
\section{Scalar, Vector, and Tensor harmonics} \label{sec:SVT}
This Appendix briefly reviews the construction of scalar, vector, and tensor harmonics on maximally symmetric three-dimensional spaces described by the metric in \eqref{eq:SFRW}. For this metric, an orthonormal triad of basis vectors can be given as follows
\begin{equation}
  \mathbf{e}_{i}=\lbrace \mathbf{e}_{r}, \mathbf{e}_{\theta}, \mathbf{e}_{\phi} \rbrace,\quad  \mathbf{e}_{r}=\sqrt{1- K r^2}\;\partial_{r},\quad \mathbf{e}_{\theta}=r^{-1}\;\partial_{\theta},\quad \mathbf{e}_{\phi}=r^{-1}\;\csc{(\theta)}\;\partial_{\phi},\label{eq:basis}
\end{equation}
together with its dual basis
\begin{equation}
    \mathbf{e}^{r}=\frac{1}{\sqrt{1- K r^2}}\;dr,\quad \mathbf{e}^{\theta}=r\;d\theta,\quad \mathbf{e}^{\phi}=r\;\sin{(\theta)}\;d\phi.\label{eq:dbasis}
\end{equation}

From the above, one can form the standard helicity vector basis $\mathbf{e}^{(m)}_{i}$ for, $m=\lbrace-1,0,1\rbrace$ as
\begin{eqnarray}
  \mathbf{e}^{(0)}_{i}&=&\mathbf{e}^{r}_{i},\quad \mathbf{e}^{\pm}_i=\frac{1}{\sqrt{2}} (\mathbf{e}^{\theta}_{i}\mp i \mathbf{e}^{\phi}_{i}),\quad \mathbf{e}^{i}_{\pm}=\frac{1}{\sqrt{2}} (\mathbf{e}^{i}_{\theta}\mp i \mathbf{e}^{i}_{\phi}),\quad \mathbf{e}^{a}_{i}=\gamma^{a}_{i},\quad a=\lbrace r,\theta,\phi\rbrace\label{eq:helbasis}
\end{eqnarray}
and the transverse-traceless (TT) tensor (spin 2) basis as
\begin{eqnarray}
\mathbf{q}^{+}_{ij}=\frac{1}{\sqrt{2}} (\mathbf{e}^{+}_{i}\mathbf{e}^{+}_{j}-\mathbf{e}^{-}_{i}\mathbf{e}^{-}_{j}),\quad \mathbf{q}^{-}_{ij}=\frac{1}{\sqrt{2}} (\mathbf{e}^{+}_{i}\mathbf{e}^{-}_{j}-\mathbf{e}^{-}_{i}\mathbf{e}^{+}_{j}),\label{eq:helTTbasis}
\end{eqnarray}
such that for a given vector $\mathbf{k}_{i}\propto \mathbf{e}^{(0)}_{i} $, along the direction of propagation of the gravitational wave, they obey the following constraints
\begin{eqnarray}
   &~& \mathbf{e}^{i}_{\pm}\mathbf{e}^{\pm}_{i}=1,\quad \mathbf{e}^{i}_{\pm}\mathbf{e}^{\mp}_{i}=0,\quad \mathbf{e}^{i}_{\pm}\mathbf{k}_{i}=\mathbf{e}_{i}^{\pm}\mathbf{k}^{i}=0,\\
   &~& \gamma^{ij}\mathbf{q}_{ij}^{{\color{red}\lambda}}=0,\quad \mathbf{k}^{i}\mathbf{q}_{ij}^{{\color{red}\lambda}}=0,\quad \mathbf{q}_{ij}^{{\color{red}\lambda}}\mathbf{q}^{ij,{\color{red}\lambda'}}=\delta^{{\color{red}\lambda},{\color{red}\lambda'}},\quad {\color{red}\lambda}=\lbrace +,- \rbrace
\end{eqnarray}

The basis vectors in \eqref{eq:helbasis} provide a convenient way for describing scalar, vector, and tensor harmonics within the 3-dimensional space. We will now elaborate on how these basis vectors facilitate the description of these harmonics.

\paragraph{\textbf{Scalar harmonics:}}
The eigenfunctions of the following differential equation
\begin{equation}
   \left( \mathcal{D}^2 +k^2\right)\Upsilon^{(n)}(\mathbf{x}) =  0,\quad k^2=n^2-K,\quad n\in \mathbb{N},\quad \begin{cases}
      \frac{n}{\sqrt{K}} >1 & \text{if }\, K>0 \\
     n\geq 0 & \text{if }\, K<0 
    \end{cases},\label{eq:scalarH}
\end{equation}
given by $\Upsilon^{(n)}$, where $\mathbf{x}$ denotes the spatial position vector, are described as the scalar spherical harmonic of order $n$ in the literature \cite{Lindblom:2017maa,Pitrou:2019ifq,Pitrou:2020lhu}. It is to be noted that in the flat space ($K = 0$), the above equation reduces to
\begin{equation}
   \left( \mathcal{D}^2 +k^2\right)\Upsilon(\mathbf{x}) =  0,\label{eq:eigenK0}
\end{equation}
whose solution in terms of spherical coordinates can be given as
\begin{equation}
    \Upsilon(\mathbf{x}) =4\pi \sum\limits_{l,m} i^{l}\,Y^{*}_{lm}(\theta_k,\phi_k)\, j_{l}(k r)\,Y_{lm}(\theta,\phi),\label{eq:solK0}
\end{equation}
which describes nothing but plane waves with $j_{l}(k r)$ being the spherical Bessel function and $ Y_{lm}(\theta,\phi)$ being the usual spherical harmonics on $\mathbb{S}^2$.

In an analogy to (\ref{eq:solK0}), the solution to eq.(\ref{eq:scalarH}) for non zero $K$ can be given as the following linear combination \cite{Lindblom:2017maa,Pitrou:2019ifq,Pitrou:2020lhu}
\begin{equation}
   \Upsilon^{(n)}(r,\theta,\phi) = \sum\limits_{l=0}^{\frac{n}{\sqrt{K}}-1}\sum\limits_{m=-l}^{l} \mathcal{A}^{n}_{lm}\Upsilon^{n}_{lm}(r,\theta,\phi)=\sum\limits_{l=0}^{\frac{n}{\sqrt{K}}-1}\sum\limits_{m=-l}^{l} \mathcal{A}^{n}_{lm}\Pi^{n}_{l}(r)Y_{lm}(\theta,\phi),
    \label{eq:Fexand}
\end{equation}
where $\mathcal{A}^{n}_{lm}$ can be understood as the Fourier coefficients, $\Upsilon^{n}_{lm}=\Pi^{n}_{l}Y_{lm}$ describes the scalar spherical harmonic of order $\lbrace n,l, m \rbrace$, and the functions $\Pi^{n}_{l}(r)$ are called as "Fock harmonics" whose definition and properties can be found in \cite{Lindblom:2017maa,Pitrou:2019ifq,Pitrou:2020lhu}. The scalar spherical harmonics $\Upsilon^{n}_{lm}$, constitute a complete and orthogonal set for the expansion of a scalar field with the following normalization conditions
\begin{equation}
    \int \frac{r^2 dr \; d\Omega} {\left(1-K r^2\right)^3} \;{\Upsilon^{*}}^{n}_{lm}(\mathbf{x})\Upsilon^{n'}_{lm}(\mathbf{x})= \begin{cases}
        \frac{\pi}{2 } \delta(n-n') \delta_{ll'} \delta_{mm'}, & if\,\,K\leq 0\\
        \frac{\pi}{2} \delta_{nn'} \delta_{ll'} \delta_{mm'}, & if\,\, K>0
    \end{cases},\label{eq:Normcond}
\end{equation}
which can be used to determine the Fourier coefficients $\mathcal{A}^{n}_{lm}$ by the following relations
\begin{equation}
 \mathcal{A}^{n}_{lm}= \begin{cases}
        \int\limits_{0}^{\infty} dn\;\int \frac{r^2 dr \; d\Omega} {\left(1-K r^2\right)^3}\;{\Upsilon^{*}}^{n}_{lm}(\mathbf{x})\Upsilon^{(n')}(\mathbf{x}), & if\,\,K\leq 0\\
         \sum\limits_{\frac{n}{\sqrt{K}}>1}^{\infty} 2\;\int \frac{r^2 dr \; d\Omega} {\left(1-K r^2\right)^3} \;{\Upsilon^{*}}^{n}_{lm}(\mathbf{x})\Upsilon^{(n')}(\mathbf{x}), & if\,\,K>0
    \end{cases}.\label{eq:coeffcond}
\end{equation} 

The scalar curvature perturbation $\Phi(\tau,\mathbf{x})$ (or any scalar function) can be expanded in the basis of scalar harmonics as
\begin{equation}
    \Phi(\tau,\mathbf{x})=\sum\limits_{\frac{n}{\sqrt{K}}>1}^{\infty} \sum\limits_{l=0}^{\frac{n}{\sqrt{K}}-1}\sum\limits_{m=-l}^{l}\;\Phi^{n}_{lm}(\tau)\;\Upsilon^{n}_{lm}(\mathbf{x}),\label{eq:Phiexpan}
\end{equation}
where the Fourier coefficients $\Phi^{n}_{lm}$, in general, depend on the conformal time and can be determined as
\begin{equation}
 \Phi^{n}_{lm}(\tau)= \begin{cases}
        \int\limits_{0}^{\infty} dn\;\int \frac{r^2 dr \; d\Omega} {\left(1-K r^2\right)^3}\;{\Upsilon^{*}}^{n}_{lm}(\mathbf{x})\;\Phi(\tau,\mathbf{x}), & if\,\,K\leq 0\\
         \sum\limits_{\frac{n}{\sqrt{K}}>1}^{\infty} \;\int \frac{r^2 dr \; d\Omega} {\left(1-K r^2\right)^3} \;{\Upsilon^{*}}^{n}_{lm}(\mathbf{x})\;\Phi(\tau,\mathbf{x}), & if\,\,K>0
    \end{cases}.\label{eq:PhiFcoeff}
\end{equation} 

\paragraph{\textbf{Vector harmonics:}}The eigenfunctions of the following differential equation
\begin{equation}
   \left( \mathcal{D}^2 +k^2-K (1-|\beta|)(2+\beta)\right)\Upsilon_{i}^{(n,\beta)}(\mathbf{x}) =  0,\quad k^2=n^2-(1+|\beta|)K,\quad \beta=\lbrace -1,0,1 \rbrace\label{eq:vectorH}
\end{equation}
given by $\Upsilon_{i}^{(n,\beta)}$ are described as the vector spherical harmonic of order $(n,\beta)$. Here, for the case $\beta=\lbrace -1,1 \rbrace$ the vector harmonics are divergence free i.e. $\mathcal{D}_{i}\Upsilon_{i}^{(n,\pm)}=0$. However, for the case $\beta=0$ the vector harmonics are not divergence free (they are longitudinal) and can be obtained from the scalar harmonics as
\begin{eqnarray}
    \Upsilon_{i}^{(n,0)}(\mathbf{x})=\sum\limits_{l\geq 1}^{\frac{n}{\sqrt{K}}-1}\sum\limits_{m=-l}^{l} \mathcal{V}^{n}_{lm}\Upsilon^{(n,0)}_{i,lm}(\mathbf{x})=\sum\limits_{l\geq 1}^{\frac{n}{\sqrt{K}}-1}\sum\limits_{m=-l}^{l} \mathcal{V}^{n}_{lm} \Pi^{n}_{i,l}(r)Y_{lm}(\theta,\phi) \mathbf{e}_i,\\ \Upsilon^{(n,0)}_{i,lm}=\frac{1}{k}\;\partial_{i} \Upsilon^{n}_{lm}(\mathbf{x}),\quad \mathcal{D}^{i}\Upsilon^{(n,0)}_{i,lm}=- k\;\Upsilon^{n}_{lm}(\mathbf{x}),\quad \left( \mathcal{D}^2 +n^2-3K \right)\Upsilon^{(n,0)}_{i,lm}(\mathbf{x}) =  0.\label{eq:DvectorH}
\end{eqnarray}
where the normalization condition for the vector harmonics described above can be given as
\begin{equation}
    \int \frac{r^2 dr \; d\Omega} {\left(1-K r^2\right)^3} \;\gamma^{ij}\,\Upsilon^{(n,0)*}_{i,lm}(\mathbf{x})\Upsilon^{i,(n',0)}_{lm}(\mathbf{x})= \begin{cases}
        \frac{\pi}{2 } \delta(n-n') \delta_{ll'} \delta_{mm'}, & if\,\,K\leq 0\\
        \frac{\pi}{2 } \delta_{nn'} \delta_{ll'} \delta_{mm'}, & if\,\, K>0
    \end{cases},\label{eq:NormcondV}
\end{equation}

Out of the three vector harmonics described here, we only require $\Upsilon_{i, lm}^{(n,0)}(\mathbf{x})$ for the computations performed in this work as using \eqref{eq:Phiexpan} and \eqref{eq:DvectorH}, one can expand the spatial derivative of the scalar curvature perturbation $\Phi(\tau,\mathbf{r})$, or any vector valued function, as
\begin{equation}
   \partial_{i} \Phi(\tau,\mathbf{x})=k \sum\limits_{\frac{n}{\sqrt{K}}>1}^{\infty}\sum\limits_{l\geq 1}^{\frac{n}{\sqrt{K}}-1}\sum\limits_{m=-l}^{l} \Phi^{n}_{lm}(\tau)\Upsilon_{i,lm}^{(n,0)}(\mathbf{x})=\mathbf{k}_i \sum\limits_{\frac{n}{\sqrt{K}}>1}^{\infty}\sum\limits_{l\geq 1}^{\frac{n}{\sqrt{K}}-1}\sum\limits_{m=-l}^{l} \Phi^{n}_{lm}(\tau)\Upsilon^{n}_{lm}(\mathbf{x}, \quad \mathbf{k}_i= k\;\mathbf{e}_i.\label{eq:Dphiexpan}
\end{equation}

\paragraph{\textbf{Tensor harmonics:}}The tensor harmonics  are described as the eigenfunctions following the differential equation
\begin{equation}
   \left( \mathcal{D}^2 +k^2-K (2-|\beta|)(3+\beta)\right)\Upsilon_{ij}^{(n,\beta)}(\mathbf{x}) =  0,\quad k^2=n^2-(1+|\beta|)K,\quad \beta=\lbrace -2, -1,0,1 ,2\rbrace\label{eq:tensorH}
\end{equation}
where depending on the values of $|\beta|$ the solutions can be split into three different classes namely
\begin{eqnarray}
    \Upsilon_{ij}^{(n,\beta)}(\mathbf{x})=\begin{cases}
        \Upsilon_{ij}^{(n,0)}, & \text{for}~|\beta|=0,~\text{Longitudinal }\\
        \Upsilon_{ij}^{(n,\pm 1)}, & \text{for}~|\beta|=1,~\text{Solonidal }\\
        \Upsilon_{ij}^{(n,\pm 2)},  & \text{for}~|\beta|=2,~ \mathcal{D}^{i}\Upsilon_{ij}^{(n,\pm 2)}=\gamma^{ij}\Upsilon_{ij}^{(n,\pm 2)}=0,~\text{Transverse traceless}
    \end{cases}.\label{eq:TensorHcases}
\end{eqnarray}

Out of the above three classes, we are only interested in the TT tensor harmonics whose expansion can be written as 
\begin{eqnarray}
    \Upsilon_{ij}^{(n,\pm 2)}(\mathbf{x})=\sum\limits_{l\geq 2}^{\frac{n}{\sqrt{K}}-1}\sum\limits_{m=-l}^{l} \mathcal{T}^{n,\pm 2}_{lm}\Upsilon^{(n,\pm 2)}_{ij,lm}(\mathbf{x})=\sum\limits_{l\geq 2}^{\frac{n}{\sqrt{K}}-1}\sum\limits_{m=-l}^{l} \mathcal{T}^{(n,\pm2)}_{lm} \Pi^{(n,\pm 2)}_{l}(r)Y_{lm}(\theta,\phi) \mathbf{q}^{\pm}_{ij},\label{eq:TTtensorH}
\end{eqnarray}
with the normalization condition for $\Upsilon^{(n,\pm 2)}_{ij,lm}$ given by
\begin{equation}
    \int \frac{r^2 dr \; d\Omega} {\left(1-K r^2\right)^3} \;\,\Upsilon^{(n,0)*}_{ij,lm}(\mathbf{x})\Upsilon^{ij,(n',0)}_{lm}(\mathbf{x})= \begin{cases}
        \frac{\pi}{2} \delta(n-n') \delta_{ll'} \delta_{mm'} \delta_{ab}, & \text{if}\,\,K\leq 0\\
        \frac{\pi}{2} \delta_{nn'} \delta_{ll'} \delta_{mm'}\delta_{ab}, & \text{if}\,\, K>0
    \end{cases},\label{eq:NormcondT}
\end{equation}

Now any TT tensor, say the second-order tensor perturbation $h^{(2)}_{ij}(\tau,\mathbf{x})$ can be expanded in terms of $\Upsilon^{(n,\pm 2)}_{ij,lm}(\mathbf{x})$ as follows
\begin{eqnarray}
   h^{(2,ss)}_{ij}(\tau,\mathbf{x})&=&\sum\limits_{\frac{n}{\sqrt{K}}>1}^{\infty}\sum\limits_{l\geq 2}^{\frac{n}{\sqrt{K}}-1}\sum\limits_{m=-l}^{l} \left(h^{(n,+)}_{lm}(\tau)\Upsilon_{ij,lm}^{(n,+2)}(\mathbf{x})+h^{(n,-)}_{lm}(\tau)\Upsilon_{ij,lm}^{(n,-2)}(\mathbf{x})\right),\nonumber\\
   &=&\sum\limits_{\frac{n}{\sqrt{K}}>1}^{\infty}\sum\limits_{l\geq 2}^{\frac{n}{\sqrt{K}}-1}\sum\limits_{m=-l}^{l} \left(h^{(n,+)}_{lm}(\tau)\Upsilon_{lm}^{(n,+2)}(\mathbf{x})\mathbf{q}^{+}_{ij}+h^{(n,-)}_{lm}(\tau)\Upsilon_{lm}^{(n,-2)}(\mathbf{x})\mathbf{q}^{-}_{ij}\right),\nonumber\\
  \label{eq:h2expansion}
\end{eqnarray}
where the functions $\Upsilon_{lm}^{(n,\pm2)}(\mathbf{x})$  are given by
\begin{equation}
   \Upsilon_{lm}^{(n,\pm2)}(\mathbf{x})= \Pi^{(n,\pm2)}_{l}(r)\;Y_{lm}(\theta,\phi).
\end{equation}
and the form of the radial functions $\Pi^{(n,\pm2)}_{l}(r)$, can be found in \cite{Abbott:1986ct}.

\section{Evaluation of the Scalar Source Term} \label{sec:SS_term}

In this Appendix, we focus on solving (\ref{eq:hij2f}),  which corresponds to the evolution equation for the second-order tensor perturbations sourced by the quadratic contributions from first-order scalar perturbations. It is important to recall here that the modes $ h_{ij}^{(2,ss)}$  correspond to the scalar-induced gravitational waves to be precise. The solution to eq.(\ref{eq:hij2f}) depends on the solutions of eq.(\ref{eq:firstscalar}) which takes a nontrivial form in the background eq.(\ref{eq:FRW}) for a non-zero spatial curvature $K$. It is obvious from the form of the Laplacian corresponding to the metric of the spatial hyper-surfaces given by eq.(\ref{eq:SFRW}) that a trivial Fourier decomposition of the field $\Phi$ in terms of plane waves is not possible and one has to use scalar harmonics described in the previous appendix. Plugging the expansion (\ref{eq:Phiexpan}) back in the eq.(\ref{eq:firstscalar}) one can get the evolution equations for the Fourier coefficients $\Phi^{n}_{lm}$ as follows
\begin{eqnarray}
    {\Phi^{n}_{lm}}''(\tau) + 3 \, \mathcal{H}\left(1 + c_{s}^{2} \right)\,{\Phi^{n}_{lm}}'(\tau) + \left(2\,\mathcal{H}' + \left(1 + 3\, c_{s}^{2}\right)\,\mathcal{H}^{2}- \left(1 + 4\, c_{s}^{2}\right) K+c_s^2 k^2\right)\,\Phi^{n}_{lm}(\tau)= 0, \label{eq:firstscalar}
\end{eqnarray}
which are easy to solve if one has the explicit form of the conformal Hubble parameter determined using from \eqref{eq:aform} as a function of the conformal time. The solution from eq.\eqref{eq:firstscalar} combined with eq.\eqref{eq:Dphiexpan} helps us to write the source term in eq.\eqref{eq:sourceSS} as 
\begin{eqnarray}
    S^{ss}_{ij}(\tau,\mathbf{x}) &=&\frac{8}{3(1+w)(\mathcal{H}^2+K)}\sum\limits_{n'l'm'}\;\sum\limits_{n''l''m''}\mathbf{k'}_i\;\mathbf{k''}_j\;\Upsilon_{l'm'}^{(n',0)}\;\Upsilon_{l''m''}^{(n'',0)} \left[ \left(\mathcal{H} \Phi^{n'}_{l'm'}(\tau)+{\Phi^{n'}_{l'm'}}'(\tau)\right)\right.\nonumber\\
    &~& \left. \left(\mathcal{H}  \Phi^{n''}_{l''m''}(\tau)+ {\Phi^{n''}_{l''m''}}'(\tau)\right) + 4 \Phi^{n'}_{l'm'}(\tau) \Phi^{n''}_{l''m''}(\tau) \right],\label{eq:sourceSS}
\end{eqnarray}

As the source term given above is transverse and traceless, one can also expand it using the TT tensor harmonics as
\begin{eqnarray}
    S^{ss}_{ij}(\tau,\mathbf{x})&=& 
    \sum\limits_{n,l,m}\left(\Upsilon_{ij,lm}^{(n,+2)}(\mathbf{x}) {S^{ss}}^{(n,+)}_{lm}(\tau)+\Upsilon_{ij,lm}^{(n,-2)}(\mathbf{x}){S^{ss}}^{(n,-)}_{lm}(\tau)\right) ,\\
    &=& \sum\limits_{n,l,m}\left(\mathbf{q}^{+}_{ij}\;\Upsilon_{lm}^{(n,+2)}(\mathbf{x})\;{S^{ss}}^{(n,+)}_{lm}(\tau)+\mathbf{q}^{-}_{ij}\;\Upsilon_{lm}^{(n,-2)}(\mathbf{x})\;{S^{ss}}^{(n,-)}_{lm}(\tau)\right),\label{eq:PTTsourceSS}
\end{eqnarray}
where the coefficients $S^{ss,n}_{ab,lm}(\tau)$ in the above expression can be determined as
\begin{equation}
    {S^{ss}}^{(n,\pm)}_{lm}(\tau)=\begin{cases}
        \int\limits_{0}^{\infty}  dn\;\int \frac{r^2 dr \; d\Omega} {\left(1-K r^2\right)^3}\;\mathbf{q}_{\pm}^{ij}\;\Upsilon_{lm}^{(n,\pm 2) *}(\mathbf{x})\;S^{ss}_{ij}(\tau,\mathbf{x}), & if\,\,K\leq 0\\
         \sum\limits_{\frac{n}{\sqrt{K}}>1}^{\infty} \;\int \frac{r^2 dr \; d\Omega} {\left(1-K r^2\right)^3} \;\mathbf{q}_{\pm}^{ij}\;\Upsilon_{lm}^{(n,\pm 2) *}(\mathbf{x})\;S^{ss}_{ij}(\tau,\mathbf{x}), & if\,\,K>0
    \end{cases}.\label{eq:FcoeffsourceSS}
\end{equation}

In order to determine the coefficients $S^{ss,n}_{lm}(\tau)$ explicitly we use the form of  $S^{ss}_{ij}(\tau,\mathbf{x})$ from \eqref{eq:sourceSS} in \eqref{eq:FcoeffsourceSS} which gives us
\begin{eqnarray}
   {S^{ss}}^{(n,\pm)}_{lm}(\tau)&=&\frac{8 }{3(1+w)(\mathcal{H}^2+K)}\sum\limits_{n'l'm'}\sum\limits_{n''l''m''}\int\limits_{0}^{\infty} dn\;\int \frac{r^2 dr \; d\Omega} {\left(1-K r^2\right)^3}\;\mathbf{q}_{\pm}^{ij}\;\mathbf{k'}_i\;\mathbf{k''}_j\;\Upsilon_{lm}^{(n,\pm 2) *}(\mathbf{x})\;\Upsilon_{l'm'}^{(n',0)}(\mathbf{x})\;\Upsilon_{l''m''}^{(n'',0)}(\mathbf{x}) \nonumber\\
    &~&\left[ \left(\mathcal{H} \Phi^{n'}_{l'm'}(\tau)+{\Phi^{n'}_{l'm'}}'(\tau)\right) \left(\mathcal{H}  \Phi^{n''}_{l''m''}(\tau)+ {\Phi^{n''}_{l''m''}}'(\tau)\right) + 4 \Phi^{n'}_{l'm'}(\tau) \Phi^{n''}_{l''m''}(\tau) \right],\nonumber \\
   &=& \frac{8 }{3(1+w)(\mathcal{H}^2+K)}\sum\limits_{n'l'm'}\;\sum\limits_{n''l''m''}\int\limits_{0}^{\infty}  dn\;\int \frac{r^2 dr \; d\Omega} {\left(1-K r^2\right)^3}\;\mathbf{q}_{\pm}^{ij}\;\mathbf{k'}_i\;\mathbf{k''}_j\;\Upsilon_{lm}^{(|n-n'|,0)}(\mathbf{x})\;\Upsilon_{l''m''}^{(n'',0)}(\mathbf{x}) \nonumber\\
    &~&\left[ \left(\mathcal{H} \Phi^{n'}_{l'm'}(\tau)+{\Phi^{n'}_{l'm'}}'(\tau)\right) \left(\mathcal{H}  \Phi^{n''}_{l''m''}(\tau)+ {\Phi^{n''}_{l''m''}}'(\tau)\right) + 4 \Phi^{n'}_{l'm'}(\tau) \Phi^{n''}_{l''m''}(\tau) \right],\nonumber \\
&=&\frac{4 }{3(1+w)(\mathcal{H}^2+K)}\sum\limits_{l'm'} \int  dn'\;k'^2 \sin^2{\theta} \cos{2\phi}\,\left[ \left(\mathcal{H} \Phi^{n'}_{l'm'}(\tau)+{\Phi^{n'}_{l'm'}}'(\tau)\right) \left(\mathcal{H}  \Phi^{|n-n'|}_{lm}(\tau)+{\Phi^{|n-n'|}_{lm}}'(\tau)\right)\right.\nonumber\\
&~&~~~~~~~~~~~~~~~~~~~~~~~~~~~~~~~\left.+ 4 \Phi^{n'}_{l'm'}(\tau) \Phi^{|n-n'|}_{lm}(\tau) \right], \label{eq:CoeffsourceSS}
\end{eqnarray}
for $K\leq 0$ and in going from the second line to the third line in the above expression we have used the normalization condition \eqref{eq:NormcondV}, converted the sum $\sum_{n'}\equiv \int dn'$ in the $n'\to \infty$ limit, and the identity $\mathbf{q}_{\pm}^{ij}\;\mathbf{k'}_i\;\mathbf{k'}_j=\frac{1}{2}k'^2 \sin^2{\theta} \cos{2\phi}$  \footnote{A similar computation can also be done for the case $K>0$ but the form of the functions ${S^{ss}}^{(n,\pm)}_{lm}$ will remain the same as given by the last line in eq.\eqref{eq:CoeffsourceSS}.}. However, going from the first line to the second line we have used the following relation
\begin{equation}
    \Upsilon_{lm}^{(n, \pm2) *}(\mathbf{x})\;\Upsilon_{l'm'}^{(n',0)}=\Upsilon_{lm}^{(|n-n'|,0)}
\end{equation}
which is a consequence of the identity given below
\begin{equation}
   \left( \mathcal{D}^2 +(k-k')^2-2K\right) \Upsilon_{lm}^{(n, \pm2) *}(\mathbf{x})\;\Upsilon_{l'm'}^{(n',0)}=\left( \mathcal{D}^2 +(k-k')^2-2K\right) \;\Upsilon_{lm}^{(|n-n'|,0)}=0
\end{equation}
implying that both the functions $\Upsilon_{lm}^{(n, \pm2) *}\;\Upsilon_{l'm'}^{(n',0)}$ and $\Upsilon_{l'm'}^{(|n-n'|,0)}$ satisfy the same differential equation.

\section{Deriving the Power spectrum of scalar induced gravitational wave}
\label{sec:AppendixD}
This Appendix gives a brief overview of obtaining the power spectrum for the scalar induced gravitational waves. Substituting the relations \eqref{eq:PTTsourceSS} and \eqref{eq:h2expansion} in \eqref{eq:hij2f} we obtain the following evolution equation
\begin{eqnarray}
    h^{(n,\pm)''}_{lm}(\tau) + 2\,\mathcal{H}\,h^{(n,\pm)'}_{lm}(\tau) +\left(n^{2} - K \right)\, h^{(n,\pm)''}_{lm}(\tau) = {S^{ss}}^{(n,\pm)}_{lm}(\tau),\label{eq:hij2t}.
\end{eqnarray}
The evolution equation \cref{eq:hij2t} admits a general solution given by:
\begin{equation}
    h^{(n,\pm)''}_{lm}(\tau) = \frac{4}{a(\tau)}\, \int^{\tau}\, d\bar{\tau}\,G^{n}_{lm}(\tau,\bar{\tau})\,a(\bar{\tau})\, {S^{ss}}^{(n,\pm)}_{lm}(\tau) \label{eq:perh2k}
\end{equation}
where $G^{n}_{lm}(\tau,\bar{\tau})$ solves the following equation:
\begin{equation}
    G^{n''}_{lm}(\tau,\bar{\tau}) + 2\,\mathcal{H}\,G^{n'}_{lm}(\tau,\bar{\tau}) + \left(n^{2} - K \right)\, G^{n}_{lm}(\tau,\bar{\tau}) = \delta(\tau,\bar{\tau}). \label{eq:green}
\end{equation}

Upon promoting the scalar field $\Phi^{n}_{lm}(\tau)$ to the quantum operator as
\begin{equation}
    \hat{\Phi}^{n}_{lm}(\tau)= \left(\frac{3 + 3\,w}{5 + 3\,w}\right)\,\zeta_{lm}^{n}\,T_{lm}^{n}(\tau) \label{eq:primordial}
\end{equation} 
and substituting the form of ${S^{ss}}^{(n,\pm)}_{lm}(\tau)$ from \eqref{eq:CoeffsourceSS} in \eqref{eq:perh2k}, one arrives at
\begin{eqnarray} 
   h^{(n,\pm)}_{lm}(\tau) &=&  4\,\left(\frac{3 + 3\,w}{5 + 3\,w}\right)^{2}\, \sum_{l'm'}\,\int dn'\,k'^2 \sin^2{\theta} \cos{2\phi}\; \zeta^{n '}_{l'm'}\,\zeta^{n-n'}_{lm}\,\int^{\tau}\, d\bar{\tau}\,G^{n}_{lm}(\tau,\bar{\tau})\,\frac{a(\bar{\tau})}{a(\tau)}\, f^{n',\abs{n-n'}}_{l'm',lm}(\tau) \nonumber \\ 
   &=& 4\,\left(\frac{3 + 3\,w}{5 + 3\,w}\right)^{2}\, \sum_{l'm'}\,\int dn'\,(n'^2 -K)\sin^2{\theta} \cos{2\phi}\; \zeta^{n'}_{l'm'}\,\zeta^{n-n'}_{lm}\, \mathcal{I}^{n',\abs{n-n'}}_{l'm',lm}(\tau) \nonumber \\
   &=& 4\,\left(\frac{3 + 3\,w}{5 + 3\,w}\right)^{2}\, \int d^{3}n'\,(n'^2 -K)\sin^2{\theta} \cos{2\phi}\; \zeta^{n'}_{l'm'}\,\zeta^{n-n'}_{lm}\, \mathcal{I}^{n',\abs{n-n'}}_{l'm',lm}(\tau) \label{eq:h_nlm}
\end{eqnarray} 
where, in the last line of equation \eqref{eq:h_nlm}, we have used $\sum_{\ell'm'} \int dn' = \int d^{3}n'$ from \cite{Pitrou:2019ifq,Pitrou:2020lhu} with the function $f^{n,\abs{n-n'}}_{l'm',lm}(\tau)$ being defined as 
\begin{equation}
    f^{n',\abs{n-n'}}_{l'm',lm}(\tau) = \left[\frac{2 }{3(1+w)(\mathcal{H}^2+K)}\, \left(\mathcal{H} \, T^{n'}_{l'm'}(\tau)+{T^{n'}_{l'm'}}'(\tau)\right) \left(\mathcal{H} \, T^{|n-n'|}_{lm}(\tau)+{T^{|n-n'|}_{lm}}'(\tau)\right) \right]+ T^{n'}_{l'm'} T^{|n-n'|}_{lm}(\tau).
\end{equation}
The two-point correlation function for induced tensor perturbations is 
\begin{equation}
    \begin{split}
        \langle h^{(n,\pm)}_{lm}(\tau) \, h^{(N,\pm)}_{LM}(\tau)  \rangle &= 16 \, \left(\frac{3 + 3\,w}{5 + 3\,w}\right)^{4} \, \int d^{3} n'd^{3}N'\,(n'^{2}-K) \,(N'^{2}-K) \sin^4{\theta} \cos^{2}{2\phi}\langle \, \zeta^{n'}_{l'm'}\,\zeta^{n-n'}_{lm}\, \zeta^{N'}_{L'M'}\,\zeta^{N-N'}_{LM} \rangle \\& \mathcal{I}^{n',\abs{n-n'}}_{l'm',lm}\mathcal{I}^{N',\abs{N-N'}}_{LM,L'M'},\label{eq:TwoPSIGW}
    \end{split}
\end{equation}
where the curvature perturbations are related to the scalar power spectrum as:
\begin{equation}
     \left\langle\hat{\Phi}^{n}_{lm}(\tau) \hat{\Phi}^{n'}_{l'm'}(\tau)\right\rangle= \left(\frac{3 + 3\,w}{5 + 3\,w}\right)^2 \left\langle\zeta_{lm}^{n}\zeta_{l'm'}^{n'}\right\rangle\; T^{n}_{lm}(\tau) T^{n'}_{l'm'}(\tau),\quad \left\langle\zeta_{lm}^{n}\zeta_{l'm'}^{n'}\right\rangle= \frac{2\pi^2}{n^3}\;\mathcal{P}(n)\;\delta(\vec{n}+\vec{n}').
\end{equation}
The scalar four-point function in \eqref{eq:TwoPSIGW} can be split into two-point functions using Wick's theorem in the following manner
     \begin{equation}
     \begin{split}
     \langle \, \zeta^{n'}_{l'm'}\,\zeta^{n-n'}_{lm}\, \zeta^{N'}_{L'M'}\,\zeta^{N-N'}_{LM} \rangle
       &=\langle \, \zeta^{n'}_{l'm'}\,  \zeta^{N'}_{L'M'} \rangle   \langle \zeta^{n-n'}_{lm} \zeta^{N-N'}_{LM} \rangle +  \langle \, \zeta^{n'}_{l'm'}\,  \zeta^{N-N'}_{LM} \rangle   \langle \zeta^{n-n'}_{lm} \zeta^{N'}_{L'M'} \rangle\\
        &=\frac{2\pi^2}{n'(n'^2-K)} \frac{2\pi^2}{\abs{n-n'}(\abs{n-n'}^2-K)} P_{\zeta}(n')  P_{\zeta}(\abs{n-n'}) \\
          &\;\;\;\;\Big(\delta(n'-N') \delta(\abs{n-n'}-\abs{N-N'}) \delta_{l'L'} \delta_{m'M'} \delta_{lL} \delta_{m M} \\
       &\;\;\;+  \delta(\abs{n-n'}-N')  \delta(n'-\abs{N-N'}) \delta_{l'L} \delta_{m'M} \delta_{lL'} \delta_{m M'}\Big).
         \end{split}
         \label{eq:wicks}
     \end{equation}

The two-point correlator in \eqref{eq:TwoPSIGW} can now be given as
\begin{equation}
\begin{split}
    \langle h^{(n,\pm)}_{lm}(\tau) \, h^{(N,\pm)}_{LM}(\tau)  \rangle &= \left(\frac{3 + 3\,w}{5 + 3\,w}\right)^{4} \, \delta(n-N)\delta_{lL}\delta_{mM}\, \int d^{3}n'\,(n'^2 -K)^{2} \frac{2\pi^2}{n'(n'^2-K)} \sin^4{\theta} \cos^{2}{2\phi}\\& \frac{2\pi^2}{\abs{n-n'}(\abs{n-n'}^2-K)}  P_{\zeta}(n')  P_{\zeta}(\abs{n-n'}) \left(\mathcal{I}^{n',\abs{n-n'}_{ss}}(\tau) \right)^{2}
    \end{split}
\end{equation}
where we have dropped the $l$ and $m$ dependence of $\mathcal{P}_{\zeta}$ and $\mathcal{I}^{n,\abs{n-n'}}_{ss}$ as we are working in isotropic Universe. It is useful to introduce the following dimensionless variables
\begin{equation}
    v = n' / n, \quad u = \abs{n-n'} / n,\quad x = \tau\,n, \quad \text{and}\quad \gamma_{n} = K / n^{2}
\end{equation}
so that the two-point correlator given above can be written as:
\begin{equation}
    \begin{split}
  \langle h^{(n,\pm)}_{lm}(\tau) \, h^{(N,\pm)}_{LM}(\tau)  \rangle &= \frac{\pi^{2}}{n^{3}} \, \left(\frac{3 + 3\,w}{5 + 3\,w}\right)^{4} \, \delta(n-N)\delta_{lL}\delta_{mM}\, \int_{0}^{\infty} du \,\int_{\abs{1-u}}^{1+u} \, dv  P_{\zeta}(n v)  P_{\zeta}(n u) \,\frac{v^{2}}{u^{2}}\,\left(1 - \frac{(1+v^{2}-u^{2})^{2}}{4 v^{2}}\right)^{2} \\& \left(1 - \frac{\gamma_{n}}{v^{2}}\right)^{3} \, \left(1 - \frac{\gamma_{n}}{u^{2}}\right)^{-1}\,\left(\mathcal{I}_{\rm ss}^{u,v}(x)\right)^{2},\label{eq:TwoPSIGWMod}
\end{split}
\end{equation}
which reduces to known form in flat limit $\gamma_{n} \rightarrow 0$\cite{Picard:2023sbz} .

\section{Evaluation of Kernel}
\label{sec:appendixE}
In this appendix, we give the details regarding the computation of the Kernel $\mathcal{I}_{\rm ss}^{u,v}(x)$ introduced in \eqref{eq:TwoPSIGWMod} and is defined as:
\begin{equation}
    \mathcal{I}_{\rm ss}^{u,v}(x) =  \int_{0}^{x}\, d \bar{x} \,f^{u,v}_{\rm ss}(\bar{x},\gamma_{n})\, \frac{a(\bar{x})}{a(x)}\,n\, G^{n}(x,\bar{x}). \label{eq:kernel1}
\end{equation}
where $G^{n}(x,\bar{x})$ is the Green's function which is solution of homogeneous equation involving $h^{n}_{\lambda}(x)$ and $f^{u,v}_{\rm ss}(\bar{x},\gamma_{n})$ takes the following form:
\begin{equation}
            f^{u,v}_{\rm ss}(\bar{x},\gamma_{n}) = T_{\Phi}^{v}(x)\,T_{\Phi}^{v}(x) + \frac{\alpha +1}{\left(\alpha +2\right)\left(\mathcal{H}^2+K\right)} \left[  \mathcal{H}^{2}\,T_{\Phi}^{v}(x)\,T_{\Phi}^{u}(x)  +  \mathcal{H}\,\left(T_{\Phi}^{u}(x)\,T'^{v}_{\Phi}(x) +\,T_{\Phi}^{v}(x)\,T'^{u}_{\Phi}(x)\right) + T'^{u}_{\Phi}(x) T'^{v}_{\Phi}(x)\right] \label{eq:fss}
\end{equation}
with $T_{\Phi}^{u}(x)$ being the scalar transfer function. We solve eqs.(\ref{eq:firstscalar}) and (\ref{eq:hij2}) to get the scalar transfer function and Green's function as
\begin{equation}
        T_{\Phi}^{u}\left(x\right) = \, 2^{\alpha +3/2} \left(c_{s}\,\,u\,\beta_{u}^{\Phi}\,x\right)^{-(\alpha +3/2)}\,\Gamma \left(\alpha + 5/2\right)\, 
   J_{\alpha + 3/2}\left(c_{s}\,\,u\,\beta_{u}^{\Phi}\,x\right) \label{eq:scalar_transfer}
\end{equation}
\begin{equation}
G^{n}\left(x, \bar{x}\right) = \frac{1}{2\,n}\,\sqrt{x\,\bar{x}} \, \left( J_{\alpha+1/2} (\bar{x}\,\beta_{n})\, Y_{\alpha+1/2} (x\, \beta_{n}) - J_{\alpha+1/2} (x\, \beta_{n})\, Y_{\alpha+1/2} (\bar{x}\, \beta_{n})\right) \label{eq:green_function}
\end{equation}
where $\beta_{u}^{\Phi} =  1 - \frac{\gamma_n}{u^{2}} \, \frac{(1 - \alpha \, (\alpha -1))}{2 c_{s}^2 \, (\alpha +1)^2 } $ and $\beta_n = 1 - \frac{\gamma_{n}\,\alpha}{6\,\left(\alpha + 1\right)}$. Inserting eqs.(\ref{eq:scalar_transfer}) and (\ref{eq:green_function}) in eq.(\ref{eq:fss}), it is straightforward is recast $f^{u,v}_{\rm ss}(\bar{x},\gamma_{n})$ in following form:
\begin{equation}
    \begin{split}
        f^{u,v}_{\rm ss}(\bar{x},\gamma_{n}) &= \frac{2^{2 \alpha + 3} \left(c_{s} x\right)^{-2 (\alpha + 1)}}{\left(\alpha + 2\right) \left(u v \beta_{u}^{\Phi} \beta_{v}^{\Phi}\right)^{\alpha + \frac{3}{2}}}  \times \Bigg[\Bigg(u \beta_{u}^{\Phi} J_{\alpha + \frac{1}{2}}\left(u c_{s} x \beta_{u}^{\Phi}\right) J_{\alpha + \frac{3}{2}}\left(v c_{s} x \beta_{v}^{\Phi}\right)  - v \beta_{v}^{\Phi} J_{\alpha + \frac{3}{2}}\left(u c_{s} x \beta_{u}^{\Phi}\right) J_{\alpha + \frac{5}{2}}\left(v c_{s} x \beta_{v}^{\Phi}\right) \\& + \frac{u v \beta_{u}^{\Phi} \beta_{v}^{\Phi} c_s x}{\alpha + 1}   J_{\alpha + \frac{5}{2}}\left(u c_{s} x \beta_{u}^{\Phi}\right) J_{\alpha + \frac{5}{2}}\left(v c_{s} x \beta_{v}^{\Phi}\right) \Bigg) + \frac{\gamma_n}{n^{2}} \Bigg( \frac{4 u x^2 \beta_{u}^{\Phi}}{3 (\alpha +1)^2} \, J_{\alpha + \frac{5}{2}}\left(u c_{s} x \beta_{u}^{\Phi}\right) J_{\alpha + \frac{3}{2}}\left(v c_{s} x \beta_{v}^{\Phi}\right) \\& -\frac{x}{(\alpha +1) c_s}  \, J_{\alpha + \frac{3}{2}}\left(u c_{s} x \beta_{u}^{\Phi}\right)  J_{\alpha + \frac{3}{2}}\left(v c_{s} x \beta_{v}^{\Phi}\right)  + \frac{4 v x^2 \beta_{v}^{\Phi}}{3 (\alpha +1)^2} \, J_{\alpha + \frac{3}{2}}\left(u c_{s} x \beta_{u}^{\Phi}\right) J_{\alpha + \frac{5}{2}}\left(v c_{s} x \beta_{v}^{\Phi}\right) \\& -\frac{5 u v x^3 c_{s} \beta_{u}^{\Phi} \beta_{v}^{\Phi}}{3 (\alpha +1)^3} \, J_{\alpha + \frac{5}{2}}\left(u c_{s} x \beta_{u}^{\Phi}\right) J_{\alpha + \frac{5}{2}}\left(v c_{s} x \beta_{v}^{\Phi}\right)\Bigg)\Bigg] \label{eq:kernel_first}
    \end{split}
\end{equation}

Now upon simplifying the $\frac{a(\bar{x})}{a(x)}\, G^{n}(x,\bar{x})$ upto leading order of $\gamma_{n}$ as:
\begin{equation}
\frac{a(\bar{x})}{a(x)}\,G^{n}(\bar{x},x) = \frac{\pi}{2}\,\frac{\bar{x}}{n} \, \left(\frac{\bar{x}}{\bar{x}}\right)^{\alpha + 1/2}\,\left[ 1 - \frac{\gamma_{n}}{6\left(1 + \alpha\right)}\,x^{2}\right]\, \Bigg[ J_{\alpha + 1/2}(\beta_{n}\,\bar{x})\,Y_{\alpha + 1/2}(\beta_{n}\,x)  + J_{\alpha + 1/2}(\beta_{n}\,x)\,Y_{\alpha + 1/2}(\beta_{n}\,\bar{x}) \Bigg] \label{eq:kernel_second}
\end{equation}
it may also be noted that while evaluating the $\mathcal{I}_{\rm ss}^{u,v}(x)$, one has to integrate the product of three Bessel functions. Using the kernel given in \eqref{eq:kernel1} and plugging in \eqref{eq:kernel_first} and \eqref{eq:kernel_second}, we can simplify the expression of the kernel upto the leading order in $\gamma_{n}$ to
\begin{equation}
        \mathcal{I}_{\rm ss}^{u,v}(x) = \frac{\pi\,2^{2 \alpha + 3} c_{s}^{-2 (\alpha +1)}\,x^{-(\alpha+1/2)}}{2\,\left(\alpha + 2\right)(2\alpha + 3) \left(u v \beta_{u}^{\Phi} \beta_{v}^{\Phi}\right)^{\alpha + \frac{1}{2}}}\,\sum_{i=1}^{4} \Big[ Y_{\alpha + 1/2} (\beta_{n}x) \, \mathcal{I}_{J}^{x,i}(u,v)  - J_{\alpha + 1/2} (\beta_{n} x) \,\mathcal{I}_{Y}^{x,i}(u,v) \Big]
\end{equation}
where we have defined $\mathcal{I}_{J,Y}^{x,i}(u,v)$ as

\begin{eqnarray}
\mathcal{I}_{J,Y}^{x,1}(x,u,v) &=& \left(1 - \frac{\gamma_{n}\,x^2}{6(\alpha+1)}\right)\int_0^{x} d\bar{x} \bar{x}^{\left(1/2 - \alpha \right)}
\left\{	
\begin{aligned}
	J_{\alpha+1/2}(\beta_{n} \bar{x})\\
	Y_{\alpha+1/2}(\beta_{n} \bar{x})
\end{aligned}
\right\}
\Bigg[ 
J_{\alpha + \frac{1}{2}}\left(u c_{s} \bar{x} \beta_{u}^{\Phi}\right) J_{\alpha + \frac{1}{2}}\left(v c_{s} \bar{x} \beta_{v}^{\Phi}\right) \nonumber\\
&& + \left(\frac{\alpha + 2}{\alpha + 1} \right) J_{\alpha + \frac{5}{2}}\left(u c_{s} \bar{x} \beta_{u}^{\Phi}\right)  J_{\alpha + \frac{5}{2}}\left(v c_{s} \bar{x} \beta_{v}^{\Phi}\right)\Bigg] \label{eq:ISS1} \\
\mathcal{I}_{J,Y}^{x,2}(x,u,v) &=& -\frac{\gamma_{n}\,\left(2 \alpha + 3\right)}{n^{2}\,c_{s}\,\left(1 + \alpha\right)^{2}\, \left(5\,\alpha + 9 \right)\,\left(u\,v\,\beta_{u}^{\Phi}\,\beta_{v}^{\Phi}\right)}\int_0^{x} d\bar{x} \bar{x}^{\left(1/2 - \alpha \right)}
\left\{	
\begin{aligned}
	J_{\alpha+1/2}(\beta_{n} \bar{x})\\
	Y_{\alpha+1/2}(\beta_{n} \bar{x})
\end{aligned}
\right\} 
\Bigg[ 
J_{\alpha + \frac{3}{2}}\left(u c_{s} \bar{x} \beta_{u}^{\Phi}\right)  \nonumber\\ && J_{\alpha + \frac{3}{2}}\left(v c_{s} \bar{x} \beta_{v}^{\Phi}\right) \Bigg] \label{eq:ISS2} \\
\mathcal{I}_{J,Y}^{x,3}(x,u,v) &=& \frac{4\,\gamma_{n}\,\left(2 \alpha + 3\right)\,\left(2 \alpha -1 \right)}{n^{2}\,c_{s}\,\left(1 + \alpha\right)\,\beta_{n}}\int_0^{x} d\bar{x} \bar{x}^{\left(3/2 - \alpha \right)}
\left\{	
\begin{aligned}
	J_{\alpha-1/2}(\beta_{n} \bar{x})\\
	Y_{\alpha-1/2}(\beta_{n} \bar{x})
\end{aligned}
\right\} 
\nonumber
\Bigg[ \frac{8\,\alpha\,\left(\alpha +2 \right)+ 3}{3 (\alpha +1)^2 (2 \alpha +3)^2}\, J_{\alpha + \frac{5}{2}}\left(u c_{s} \bar{x} \beta_{u}^{\Phi}\right) J_{\alpha + \frac{5}{2}}\left(v c_{s} \bar{x} \beta_{v}^{\Phi}\right)  \\&& +\frac{4\left(\alpha + 2\right)}{(\alpha +1) (2 \alpha +3)^2} \, J_{\alpha + \frac{1}{2}}\left(u c_{s} \bar{x} \beta_{u}^{\Phi}\right) J_{\alpha + \frac{1}{2}}\left(v c_{s} \bar{x}\beta_{v}^{\Phi}\right) \Bigg] \label{eq:ISS3} \\
\mathcal{I}_{J,Y}^{x,3}(x,u,v) &=& \frac{4\,\gamma_{n}\,\left(2 \alpha + 3\right)\,\left(2 \alpha -1 \right)}{n^{2}\,c_{s}\,\left(1 + \alpha\right)\,\beta_{n}}\int_0^{x} d\bar{x} \bar{x}^{\left(5/2 - \alpha \right)}
\left\{	
\begin{aligned}
	J_{\alpha-3/2}(\beta_{n} \bar{x})\\
	Y_{\alpha-3/2}(\beta_{n} \bar{x})
\end{aligned}
\right\} \nonumber
\Bigg[\frac{8\,\alpha\,\left(\alpha +2 \right)+ 3}{3 (\alpha +1)^2 (2 \alpha +3)^2} \, J_{\alpha + \frac{5}{2}}\left(u c_{s} \bar{x} \beta_{u}^{\Phi}\right) J_{\alpha + \frac{5}{2}}\left(v c_{s} \bar{x} \beta_{v}^{\Phi}\right)  \\&& +\frac{4\left(\alpha + 2\right)}{(\alpha +1) (2 \alpha +3)^2}\, J_{\alpha + \frac{1}{2}}\left(u c_{s} \bar{x} \beta_{u}^{\Phi}\right) J_{\alpha + \frac{1}{2}}\left(v c_{s} \bar{x}\beta_{v}^{\Phi}\right) \Bigg]\label{eq:ISS4} 
\end{eqnarray}
It is not possible to compute the integrals for general values of $x$ but one can perform the the analytical solution in sub-horizon and super-horizon limits. In this work, we are only interested in scales those are deep inside horizon before reheating so all scales are in sub-horizon limit i.e. $x \equiv n\,\tau \gg 1$. In such scenario, we can approximate above integrals by sending their upper limit from $x$ to $\infty$ as prescribed in \cite{threebesselI,originalbessel,NIST:DLMF}. Now, in order to further simplify the kernel we use the following formula as elucidated in \cite{threebesselI} such that for $|a-b|<c<a+b$,
\begin{align}
\int_0^{\infty} d\tilde x \tilde x^{1-\beta}
\left\{	
\begin{aligned}
	J_\beta(c\tilde x)\\
	Y_\beta(c\tilde x)
\end{aligned}
\right\}
J_{\nu}(a\tilde x)J_{\nu}(b\tilde x)=\frac{1}{\pi}\sqrt{\frac{2}{\pi}}\frac{(ab)^{\beta-1}}{c^\beta}\left(\sin\varphi\right)^{\beta-1/2}\left\{	
\begin{aligned}
	\frac{\pi}{2}\mathsf{P}^{-\beta+1/2}_{\nu-1/2}(\cos\varphi)\\
	-\mathsf{Q}^{-\beta+1/2}_{\nu-1/2}(\cos\varphi)
\end{aligned}
\right\} \label{eq:besslint}
\end{align}
where
\begin{align}
16\Delta^2\equiv\left(c^2-(a-b)^2\right)\left((a+b)^2-c^2\right)\quad,\quad
\cos\varphi=\frac{a^2+b^2-c^2}{2ab}\quad,\quad\sin\varphi=\frac{2\Delta}{ab}\,.
\end{align}
and for $c>a+b$
\begin{align}
\int_0^{\infty} d\tilde x \tilde x^{1-\beta}
&\left\{	
\begin{aligned}
	J_\beta(c\tilde x)\\
	Y_\beta(c\tilde x)
\end{aligned}
\right\}
J_{\nu}(a\tilde x)J_{\nu}(b\tilde x)\nonumber\\&
=\frac{1}{\pi}\sqrt{\frac{2}{\pi}}\frac{(ab)^{\beta-1}}{c^\beta}\left(\sinh\phi\right)^{\beta-1/2}\Gamma[\nu-\beta+1]{\cal Q}^{-\beta+1/2}_{\nu-1/2}(\cosh\phi)\left\{	
\begin{aligned}
	-\sin\left[(\nu-\beta)\pi\right]\\
	\cos\left[(\nu-\beta)\pi\right]
\end{aligned}
\right\}
\end{align}
\begin{align}
16\tilde\Delta^2\equiv\left(c^2-(a-b)^2\right)\left(c^2-(a+b)^2\right) \quad {,}\quad 
\cosh\phi=\frac{c^2-(a^2+b^2)}{2ab}\quad,\quad\sinh\phi=\frac{2\tilde\Delta}{ab}\,.
\end{align}

In the above expressions $\mathsf{P}_{\mu}^{\nu}(x)$ and $\mathsf{Q}_{\mu}^{\nu}(x)$  are the Legendre functions on the cut (or Ferrers Functions) while $\mathcal{Q}_{\mu}^{\nu}(x)$ is the associated Legendre polynomials of second kind whose definitions can be found in NIST database \cite{NIST:DLMF}. Using the above formulae for the integral, we obtain the kernel in following form:
\begin{align}
    \mathcal{I}_{\rm ss}^{u,v}(x) &= \frac{2^{\alpha+3}\,x^{-(\alpha+1)}\, c_s^{-3}}{\pi \, \left(\alpha + 2\right)\,\left(2 \alpha + 3\right)\,\left(u v \beta_{u}^{\Phi} \beta_{v}^{\Phi}\, \beta_n \right)^{\alpha + 1}} \Bigg\{ \frac{\pi}{2} \, Y_{\alpha+1/2}\left(\beta_{n}\, x\right) 
    \Bigg[\left(1 - \frac{\gamma_{n}}{6(\alpha + 1)}\,x^{2}\right) \tilde{\mathcal{I}}^{1}_{J}(u,v) \nonumber \\ 
    &\quad +  \gamma_{n}\, \frac{2 \,(2 \alpha + 3)}{c_s\, n^{2} \, \pi \, (1+\alpha)^{2}\, (5 \alpha + 9 )\, u v \beta_{u}^{\Phi} \beta_{v}^{\Phi}} \, \tilde{\mathcal{I}}^{2}_{J}(u,v) +  \gamma_{n} \, \frac{4\,\left(2 \alpha -1 \right)\,(3 + 2\alpha)}{n^{2} \,\, \pi \,c_s \,(1 + \alpha)} \Big[ \, \tilde{\mathcal{I}}^{3}_{J}(u,v)  + 2\,\frac{\beta_n}{c_s^{2}} \,  \tilde{\mathcal{I}}^{4}_{J}(u,v)\Big] \Bigg] \nonumber \\ 
    &\quad + \, J_{\alpha+1/2}\left(\beta_{n}\, x\right) \times \Bigg[\left(1 - \frac{\gamma_{n}}{6(\alpha + 1)}\,x^{2}\right) \tilde{\mathcal{I}}^{1}_{Y}(u,v) +  \gamma_{n}\, \frac{2 \,(2 \alpha + 3)}{c_s\, n^{2} \, \pi \, (1+\alpha)^{2}\, (5 \alpha + 9 )\, u v \beta_{u}^{\Phi} \beta_{v}^{\Phi}} \, \tilde{\mathcal{I}}^{2}_{Y}(u,v) \nonumber \\ 
    &\quad +  \gamma_{n} \, \frac{4\,\left(2 \alpha -1 \right)\,(3 + 2\alpha)}{n^{2} \,\, \pi \,c_s \,(1 + \alpha)} \Big[ \, \tilde{\mathcal{I}}^{3}_{Y}(u,v)  + 2\,\frac{\beta_n}{c_s^{2}} \,  \tilde{\mathcal{I}}^{4}_{Y}(u,v)\Big] \Bigg]\Bigg\}  \nonumber
\end{align}
\begin{align}
    &= \frac{2^{\alpha+3}\,\left(u \,v \,\beta_{u}^{\Phi} \,\beta_{v}^{\Phi}\, \beta_n \,x\right)^{-(\alpha+1)}}{\pi \, \, c_s^{-3}\,\left(\alpha + 2\right)\,\left(2 \alpha + 3\right)} \, \times \Bigg\{ \frac{\pi}{2} \, Y_{\alpha+ \frac{1}{2}}\left(\beta_{n}\, x\right) \, \mathcal{I}_{J}(u,v)  + J_{\alpha+ \frac{1}{2}}\left(\beta_{n}\, x\right) \, \mathcal{I}_{Y}(u,v) \Bigg\}
\end{align}
with the definitions of $\tilde{\mathcal{I}^{i}}_{J/Y}$ given as follows
\begin{equation}
    \begin{aligned}
        \mathcal{\tilde{I}}^{1}_{J}(u,v) &= Z^{\alpha} \,\Big[\mathbf{P}_{\alpha}^{-\alpha}(s) + \left(\frac{\alpha + 2}{\alpha + 1} \right) \, \mathbf{P}_{\alpha + 2}^{-\alpha}(s)\Big]\, \Theta(\abs{s}-1)\,, \\
        \mathcal{\tilde{I}}^{1}_{Y}(u,v) &= \Big\{ Z^{\alpha}\,\Big[\mathbf{Q}_{\alpha}^{-\alpha}(s) + \left(\frac{\alpha + 2}{\alpha + 1} \right) \, \mathbf{Q}_{\alpha + 2}^{-\alpha}(s)\Big]\,  \Theta(\abs{s}-1)  - \tilde{Z}^{\alpha} \Big[\mathcal{Q}_{\alpha}^{-\alpha}(\tilde{s}) + \left(\frac{\alpha + 2}{\alpha + 1} \right) \, \mathcal{Q}_{\alpha + 2}^{-\alpha}(\tilde{s})\Big]\,\Theta( 1 - \abs{s}) \Big\}\,, \\
        \mathcal{\tilde{I}}^{2}_{J}(u,v) &= Z^{\alpha}\,\mathbf{P}^{-\alpha}_{\alpha+1}(s) \, \Theta (\abs{s}-1)\,, \\
        \mathcal{\tilde{I}}^{2}_{Y}(u,v) &=  Z^{\alpha}\,\mathbf{Q}_{\alpha+1}^{-\alpha}(s) \,  \Theta(\abs{s}-1) - \tilde{Z}^{\alpha} \mathcal{Q}_{\alpha+1}^{-\alpha}(\tilde{s}) \, \Theta( 1-\abs{s} ) \,, \\
        \mathcal{\tilde{I}}^{3}_{J}(u,v) &= Z^{\alpha -1}\, \Bigg[ A(\alpha)  \, \mathbf{P}^{-\alpha + 1}_{\alpha + 2}(s) + B(\alpha) \, \mathbf{P}^{-\alpha + 1}_{\alpha} \Bigg] \, \Theta (\abs{s}-1) \,, \\
        \mathcal{\tilde{I}}^{3}_{Y}(u,v) &= \Big\{ Z^{\alpha-1}\,\Big[ A(\alpha)\,\mathbf{Q}_{\alpha+2}^{1-\alpha}(s)  + B(\alpha) \, \mathbf{Q}_{\alpha}^{1-\alpha}(s)\Big]\,  \Theta(\abs{s}-1)  - \tilde{Z}^{\alpha-1} \Big[ A(\alpha)\,\mathcal{Q}_{\alpha+2}^{1-\alpha}(\tilde{s}) - 2\, B(\alpha) \, \mathcal{Q}_{\alpha}^{1-\alpha}(\tilde{s})\Big]\, \Theta(1-\abs{s}) \Big\} \,, \\
        \mathcal{\tilde{I}}^{4}_{J}(u,v) &= Z^{\alpha -2}\, \Bigg[A(\alpha) \, \mathbf{P}^{-\alpha + 2}_{\alpha + 2}(s)  + B(\alpha) \, \mathbf{P}^{-\alpha + 2}_{\alpha} \Bigg] \, \Theta (\abs{s}-1) \,, \\
        \mathcal{\tilde{I}}^{4}_{Y}(u,v) &= \Big\{ Z^{\alpha-2}\,\Big[ A(\alpha) \,\mathbf{Q}_{\alpha+2}^{2-\alpha}(s)  + B(\alpha) \, \mathbf{Q}_{\alpha}^{2-\alpha}(s)\Big]\,  \Theta(\abs{s}-1)  - \tilde{Z}^{\alpha-2} \Big[ A(\alpha) \,\mathcal{Q}_{\alpha+2}^{2-\alpha}(\tilde{s})  + 2\,B(\alpha)\, \mathcal{Q}_{\alpha}^{2-\alpha}(\tilde{s})\Big]\, \Theta(1-\abs{s}) \Big\}\,.
    \end{aligned}
\end{equation}
where in the above, we have the further introduced the following parameters
\begin{eqnarray}
    s &=& \frac{(u \beta_{u}^{\Phi})^{2} + (v \beta_{v}^{\Phi})^{2}  - \beta_{n}^{2} cs^{-2}}{2 u v \beta_{u}^{\Phi} \beta_{v}^{\Phi}}, \qquad Z^{2} =  (2 u \,v \beta_{u}^{\Phi} \, \beta_{v}^{\Phi})^{2} \left(1-s^{2} \right) \nonumber \\
    \tilde{Z}^{2} &=& -Z^{2} ,\qquad  A(\alpha) =  \frac{8\,\alpha\,\left(\alpha +2 \right)+ 3}{3 (\alpha +1)^2 (2 \alpha +3)^2} \qquad \, \text{and}\qquad B(\alpha) =  \frac{4\left(\alpha + 2\right)}{(\alpha +1) (2 \alpha +3)^2}
\end{eqnarray}
for brevity and compactness of the expressions.